\documentclass[twocolumn,showpacs,preprintnumbers,amsmath,amssymb,pra,aps,
    superscriptaddress,longbibliography]{revtex4-1}
\usepackage{graphicx}
\usepackage{multirow}
\usepackage{amsmath}

\usepackage{soul}
\usepackage{verbatim}
\usepackage{bm}
\usepackage{outlines}
\usepackage{siunitx}
\usepackage[space]{grffile}
\usepackage[dvipsnames]{xcolor}
\usepackage{xr}
\usepackage{color}
\usepackage{tikz}
\usepackage{mathtools}
\usepackage{amssymb}

\newcommand\myshade{85}
\colorlet{mylinkcolor}{violet}
\colorlet{mycitecolor}{YellowOrange}
\colorlet{myurlcolor}{Aquamarine}
\usepackage{hyperref}
\hypersetup{
  linkcolor  = mylinkcolor!\myshade!black,
  citecolor  = mycitecolor!\myshade!black,
  urlcolor   = myurlcolor!\myshade!black,
  colorlinks = true,
}
\usepackage[capitalize]{cleveref}

\usetikzlibrary{calc, positioning}

\newcommand{\MHz}{\mathrm{MHz}}
\newcommand{\GHz}{\mathrm{GHz}}

\newcommand{\us}{\mu\mathrm{s}}
\newcommand{\ns}{\mathrm{ns}}

\newcommand{\ket}[1]{\left\lvert #1 \right\rangle}
\newcommand{\bra}[1]{\left\langle #1 \right\rvert}
\newcommand{\braket}[2]{\langle #1| #2\rangle}

\newcommand{\LSQ}{\QM}
\newcommand{\MSQ}{\QH}
\newcommand{\LorMSQ}{q_i}

\newcommand{\condphase}{\phi_{2Q}}
\newcommand{\CZ}{\mathrm{CZ}} 

\newcommand{\TCZ}{T_{\mathrm{CZ}}}
\newcommand{\TtwoQ}{T_{2Q}}
\newcommand{\ToneQ}{T_{1Q}}

\newcommand{\QH}{q_\mathrm{H}}
\newcommand{\QM}{q_\mathrm{M}}
\newcommand{\QL}{q_\mathrm{L}}
\newcommand{\Qtarget}{q_\mathrm{targ.}}
\newcommand{\Qcontrol}{q_\mathrm{contr.}}

\newcommand{\tr}{\text{Tr}}

\newcommand{\infid}{\varepsilon} 
\newcommand{\fid}{F}

\newcommand{\leak}{L_1}
\newcommand{\estleak}{\widetilde{L_1}}

\newcommand{\infidClifford}{\infid^{\mathrm{Cl.}}}
\newcommand{\infidCZ}{\infid^{\CZ}}
\newcommand{\leakClifford}{\leak^{\mathrm{Cl.}}}
\newcommand{\leakCZ}{\leak^{\CZ}}
\newcommand{\infidInterleaved}{\infid^{\mathrm{Int.}}}
\newcommand{\leakInterleaved}{\leak^{\mathrm{Int.}}}
\newcommand{\seepClifford}{\seep^{\mathrm{Cl.}}}

\newcommand{\seep}{L_2}
\newcommand{\compsub}{\mathcal{X}_1}
\newcommand{\leaksub}{\mathcal{X}_2}

\newcommand{\netzero}{\mathrm{NZ}}

\newcommand{\missingfrac}{m}

\newcommand{\BSphaseNZ}{\condphase^\mathrm{NZ}}
\newcommand{\leakNZ}{\leak^\mathrm{NZ}}
\newcommand{\BSphaseHalf}{\condphase^\mathrm{half}}
\newcommand{\leakHalf}{\leak^\mathrm{half}}
\newcommand{\PSphase}{\varphi}

\newcommand{\thetaf}{\theta_f}
\newcommand{\detuning}{\epsilon}
\newcommand{\qubitHamiltonian}{H_\mathrm{subspace}}
\newcommand{\anharmonicity}{\eta}
\newcommand{\PhiT}{\Phi_{\mathrm{target}}}
\newcommand{\targetFlux}{\PhiT} 

\newcommand{\VAWG}{V_{\mathrm{AWG}}}
\newcommand{\actualFlux}{\Phi}

\newcommand{\impulseresponse}{h}
\newcommand{\transferfunc}{\mathcal{H}} 

\newcommand{\echo}{\mathrm{E}}
\newcommand{\ramsey}{*}

\newcommand{\Tone}{T_{1}}
\newcommand{\Ttwos}{T_{2}^*}

\newcommand{\abs}[1]{\left| #1 \right|}

\begin{document}
\title{A fast, low-leakage, high-fidelity two-qubit gate for a programmable superconducting quantum computer}

\newcommand{\QuTech}{\affiliation{QuTech, Delft University of Technology, P.O. Box 5046, 2600 GA Delft, The Netherlands}}
\newcommand{\Kavli}{\affiliation{Kavli Institute of Nanoscience, Delft University of Technology, P.O. Box 5046, 2600 GA Delft, The Netherlands}}
\newcommand{\JARA}{\affiliation{JARA Institute for Quantum Information, Forschungszentrum Juelich, D-52425 Juelich, Germany}}
\newcommand{\TNO}{\affiliation{Netherlands Organisation for Applied Scientiﬁc Research (TNO), P.O. Box 96864, 2509 JG The Hague, The Netherlands}}

\author{M.~A.~Rol}\QuTech\Kavli
\author{F. Battistel}\QuTech
\author{F.~K.~Malinowski}\QuTech\Kavli
\author{C.~C.~Bultink}\QuTech\Kavli
\author{B.~M.~Tarasinski}\QuTech\Kavli
\author{R.~Vollmer}\QuTech\Kavli
\author{N.~Haider}\QuTech\TNO
\author{N.~Muthusubramanian}\QuTech\Kavli
\author{A.~Bruno}\QuTech\Kavli
\author{B.~M.~Terhal}\QuTech\JARA
\author{L.~DiCarlo}\QuTech\Kavli
\date{\today}

\begin{abstract}
A common approach to realize conditional-phase (CZ) gates in transmon qubits relies on flux control of the qubit frequency to make computational states interact with non-computational ones using a fast-adiabatic trajectory to minimize leakage.
We develop a bipolar flux-pulsing method with two key advantages over the traditional unipolar variant.
First, the action of the bipolar pulse is robust to long-timescale linear-dynamical distortions in the flux-control line, facilitating tuneup and ensuring atomic repeatability.
Second, the flux symmetry of the transmon Hamiltonian makes the conditional phase and the single-qubit phase of the pulsed qubit first-order insensitive to low-frequency flux noise, increasing fidelity.
By harnessing destructive interference to minimize leakage, the bipolar pulse can approach the speed limit set by the exchange coupling.
We demonstrate a repeatable, high-fidelity ($99.1\%$), low-leakage ($0.1\%$), and fast ($40~\ns$) CZ gate in a circuit QED quantum processor.
Detailed numerical simulations with excellent match to experiment show that leakage is dominated by remaining short-timescale distortions and fidelity is limited by high-frequency flux noise.
\end{abstract}

\maketitle
A steady increase in qubit counts~\cite{Otterbach17,Knight17,Kelly18,Intel18} and operation fidelities~\cite{Barends14,Rol16,Hong19,Sheldon16b,Heinsoo18} allows quantum computing platforms using monolithic superconducting quantum hardware to target outstanding challenges such as quantum advantage~\cite{Boixo18,Neill18,Bravyi18}, quantum error correction (QEC)~\cite{Kelly15,Riste15,Takita16,Bultink19_ZZXX,Kraglund19}, and quantum fault tolerance (QFT)~\cite{Fowler12,Martinis15}. All of these pursuits require two-qubit gates with fidelities exceeding $99\%$, fueling very active research.

There are three main types of two-qubit gates in use for transmon qubits, all of which harness exchange interactions between computational states ($\ket{ij}, i,j \in \{0,1\}$)
 or between computational  and non-computational states ($i$ or $j\geq 2$), mediated by a coupling bus or capacitor.
Cross-resonance gates~\cite{Chow11,Sheldon16b} exploit the exchange interaction between $\ket{01}$ and $\ket{10}$ using microwave-frequency transversal drives.
Parametric gates~\cite{Caldwell18,Hong19} employ radio-frequency longitudinal drives, specifically flux pulses modulating the qubit frequency, to generate sidebands of resonance between $\ket{01}$ and $\ket{10}$ for iSWAP or between $\ket{11}$ and $\ket{02}$ or $\ket{20}$ for conditional-phase (CZ).
The oldest approach~\cite{Strauch03,DiCarlo09} uses baseband flux pulses to tune $\ket{11}$ into near resonance with $\ket{02}$ to realize CZ.
Either because they explicitly use non-computational states, or because of frequency crowding and the weak transmon anharmonicity, the three approaches are vulnerable to leakage of information from the computational space.
Leakage is very problematic in applications such as QEC, complicating the design of error decoders and/or demanding operational overhead to generate seepage~\cite{Aliferis07,Ghosh13_B,Fowler13,Suchara15,Ghosh15}, generally reducing the error thresholds for QFT.
This threat has motivated the design of fast-adiabatic pulsing methods~\cite{Martinis14} to mitigate leakage and architectural choices in qubit frequency and coupler arrangements~\cite{Versluis17} to explicitly avoid it.
Surprisingly, many recent demonstrations~\cite{Sheldon16b,Wang18,Hong19} of two-qubit gates place emphasis on reaching or approaching $99\%$ fidelity without separately quantifying leakage.

Although baseband flux pulsing produces the fastest two-qubit gates to date ($30-45~\ns$), two challenges have kept it from becoming the de facto two-qubit gating method. First, because the pulse displaces one qubit $0.5-1~\GHz$ below its flux-symmetry point, i.e., the sweetspot, the temporary first-order sensitivity to flux noise increases dephasing and impacts fidelity. The second challenge is non-atomicity. If uncompensated, linear-dynamical distortions in the flux-control lines originating from limited waveform-generator bandwidth, high-pass bias tees, low-pass filters, impedance mismatches, on-chip response, etc., can make the action of a pulse depend on the detailed history of flux pulses applied. To date, predistortion corrections have been calculated in advance, requiring prior knowledge of the timing of all the flux-pulse-based operations required by the quantum circuit, and significant waveform memory. This standard practice is incompatible with real-time determination and execution of operations, as is required for control flow and feedback in a fully programmable quantum computer~\cite{Fu17,Fu19}.

In this Letter, we introduce a fast ($40~\ns$),  low-leakage ($0.1\%$), high-fidelity ($99.1\%$), and repeatable flux-pulse-based $\CZ$ gate suitable for a full-stack quantum computer executing operations in real time on transmon-based quantum hardware.
These attractive characteristics are enabled by a zero-average bipolar flux-pulsing method, nicknamed Net-Zero (NZ), which uses the $\ket{11}\leftrightarrow\ket{02}$ avoided crossing twice. Harnessing the analogy to a Mach-Zehnder interferometer, NZ exploits destructive interference to minimize leakage to $\ket{02}$ while approaching the speed limit set by the
exchange coupling in the two-excitation manifold.
The flux symmetry of the transmon Hamiltonian makes the conditional phase and the single-qubit phase acquired by the pulsed qubit first-order insensitive to low-frequency flux noise, increasing fidelity relative to a unipolar pulse.
Crucially, the zero-average characteristic makes NZ insensitive to long timescale linear-dynamical distortions remaining in the flux-control line after best efforts at real-time pre-compensation, making the $\CZ$ gate repeatable.
Detailed numerical simulations supplied with calibrated experimental parameters and  direct measurement of remaining short-timescale distortions show an excellent match to experiment, and indicate that fidelity is limited by high-frequency flux noise while leakage is dominated by the remaining short-timescale distortions.

\begin{figure}
  \centering
    \includegraphics{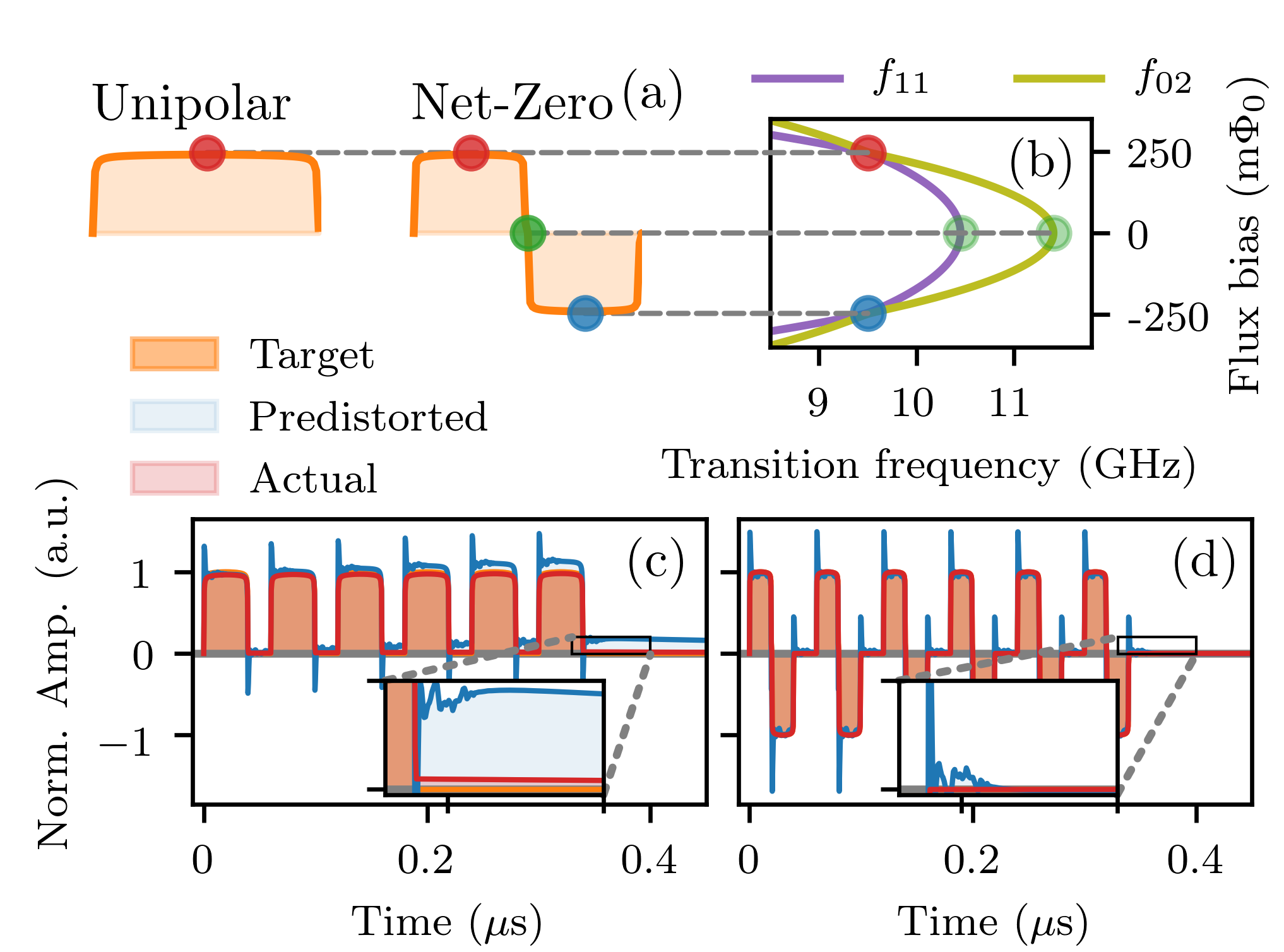}
  \caption{\label{fig:concept_net_zero}(a) Schematic representation of unipolar and $\netzero$ flux pulses that tune into resonance with (b) the $\ket{11}\leftrightarrow \ket{02}$ avoided crossing in order to perform $\CZ$ gates.
  Repeated applications of unipolar (c) and $\netzero$ (d) CZ pulses showing the target (orange), predistorted (blue), and actual (red) waveforms for an imperfect distortion correction.
  The insets in (c) and (d) show the differing accumulation in the required predistortion correction.
  }
\end{figure}

The ideal $\CZ$ gate is described by the transformation:
\begin{equation}
U = \begin{pmatrix}
1  &  0  &   0   &  0\\
0  &  e^{i\phi_{01}}  &   0   &  0\\
0  &  0  &   e^{i\phi_{10}}   &  0\\
0  &  0  &   0   &  e^{i\phi_{11}}
\end{pmatrix},
\label{eq:CZ_matrix}
\end{equation}
in the computational basis $\{\ket{00}, \ket{01}, \ket{10}, \ket{11}\}$, where the single-qubit phases $\phi_{01}$ and $\phi_{10}$ are even multiples of $\pi$ and the conditional phase defined by $\condphase=\phi_{11}-\phi_{01}-\phi_{10}$ is an odd multiple of $\pi$.
A $\CZ$ gate of total duration $\TCZ = \TtwoQ + \ToneQ$ can be realized in two steps.
First, a strong flux pulse on the higher frequency qubit moves $\ket{11}$ into the avoided crossing with $\ket{02}$ and back to acquire $\condphase$.
Next, simultaneous weaker pulses on both qubits adjust the single-qubit phases.
We compare two types of flux pulses, the (unipolar) pulse introduced in~\cite{Martinis14} and the $\netzero$ pulse [\cref{fig:concept_net_zero}(a)].
The $\netzero$ pulse consists of two back-to-back unipolar pulses of half the duration and opposite amplitude.
Experiments are performed on a pair of flux-tunable transmons described in more detail in the Supplemental Material~\cite{Suppmaterial}.

Because of distortions, the waveform $\VAWG(t)$ specified in an arbitrary waveform generator (AWG) does not result in the qubit experiencing the targeted flux $\PhiT(t)$.
These distortions can be described as a linear time-invariant system that transduces voltage to flux and is characterized by its impulse response $\impulseresponse(t)$.
It is possible to measure $\impulseresponse(t)$ at the qubit using our Cryoscope technique~\cite{Rol19_cryoscope} and construct an inverse filter $\tilde{\impulseresponse}^{-1}$ to correct for distortions.
By performing a convolution of the desired signal $\PhiT(t)$ with $\tilde{\impulseresponse}^{-1}$, known as a predistortion correction, the qubit experiences the pulse
\begin{equation}
\label{eq:convolution}
\Phi(t) = \impulseresponse \ast \VAWG(t) = \impulseresponse \ast (\tilde{\impulseresponse}^{-1} \ast \PhiT )(t).
\end{equation}
The predistortion corrections are performed using a combination of real-time filters implemented in a Zurich Instruments HDAWG and a short ($20~\ns$) FIR filter implemented off line.

By eliminating the DC component of the pulse, $\netzero$ $\CZ$ gates are resilient to remaining long-timescale distortions~\cite{Johnson11}.
Because the transmon Hamiltonian is symmetric with respect to the sweetspot, it is possible to use both positive and negative amplitudes to perform a $\CZ$ gate~[\cref{fig:concept_net_zero}(b)]
while also satisfying the zero-average condition
\begin{equation}
\label{eq:netzero_cond}
\int_0^{\TCZ} \PhiT(t') dt' = 0.
\end{equation}
If \cref{eq:netzero_cond} holds, the DC component is zero and the components in the Fourier transform $\PhiT(\omega)$ at frequencies $\omega \lesssim \frac{2\pi}{\TCZ}$ are suppressed.
Writing \cref{eq:convolution} in the Fourier domain: $\Phi(\omega) =\transferfunc(\omega) \cdot \tilde{\transferfunc}^{-1}(\omega)\cdot\PhiT(\omega)$,
it follows that if $\PhiT(\omega)$ does not contain any components at $\omega<\frac{2\pi}{\TCZ}$, then $\Phi(\omega)$ does not depend on any components of $\transferfunc(\omega)$ at frequencies $\omega<\frac{2\pi}{\TCZ}$.
As a consequence, the required corrections for $\netzero$ pulses do not accumulate, eliminating the need for accurate long-timescale distortion corrections and the resulting history-dependent errors~[\cref{fig:concept_net_zero}(d)].

\begin{figure}
  \centering
    \includegraphics{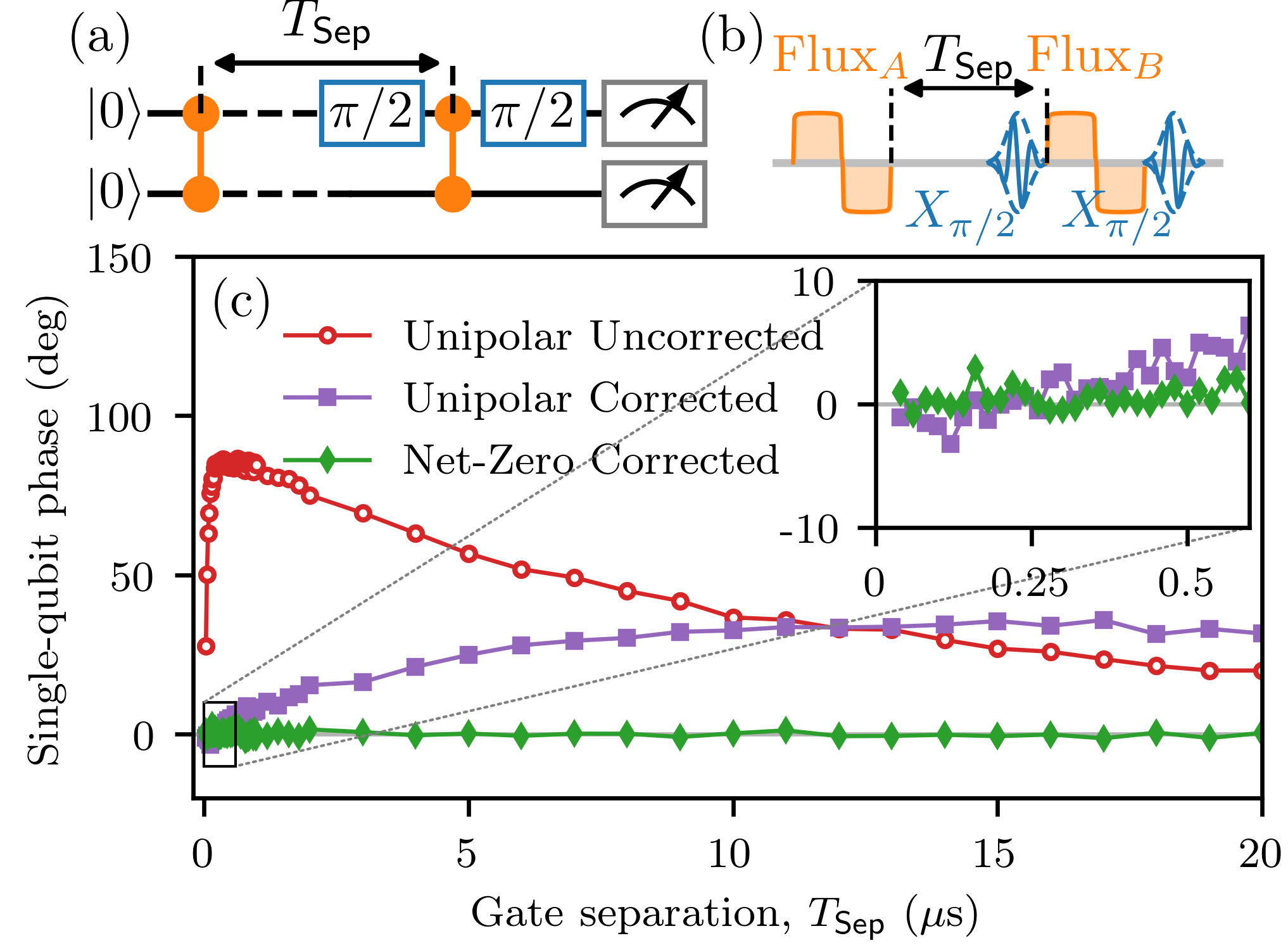}
    \caption{\label{fig:repeatability}
    History dependence of flux pulses.
    (a) Circuit used to measure the phase acquired during a flux pulse as a function of the separation time $T_\mathrm{Sep}$ to another flux pulse earlier in time and (b) the corresponding pulses.
    Both pulses are calibrated to correspond to a $\CZ$ gate.
    (c) Acquired single-qubit phase for unipolar pulses without (red), and with (purple) predistortion corrections and $\netzero$ pulses with predistortion corrections (green).
  }
\end{figure}

To measure the repeatability of $\CZ$ gates, the phase ($\phi_{01}$) acquired by the pulsed qubit during a $\CZ$ gate is measured as a function of the separation time $T_\mathrm{Sep}$ to an earlier $\CZ$ gate~(\cref{fig:repeatability}).
Because of the detuning from the sweetspot, a small change in amplitude during the pulse leads to a significant change in frequency.
This makes the phase acquired during the pulse very sensitive to distortions.
We observe that not correcting distortions leads to significant phase errors ($\sim 80~\deg$).
Correcting distortions using a predistortion filter keeps the error small ($<10~\deg$) for the first $500~\ns$ but shows clear history-dependent behavior for longer timescales.
Using $\netzero$ pulses in combination with a predistortion filter eliminates all history dependence.
Hence, we conclude that $\netzero$ pulses are robust against remaining long-timescale distortions.

\begin{figure}
  \centering
    \includegraphics{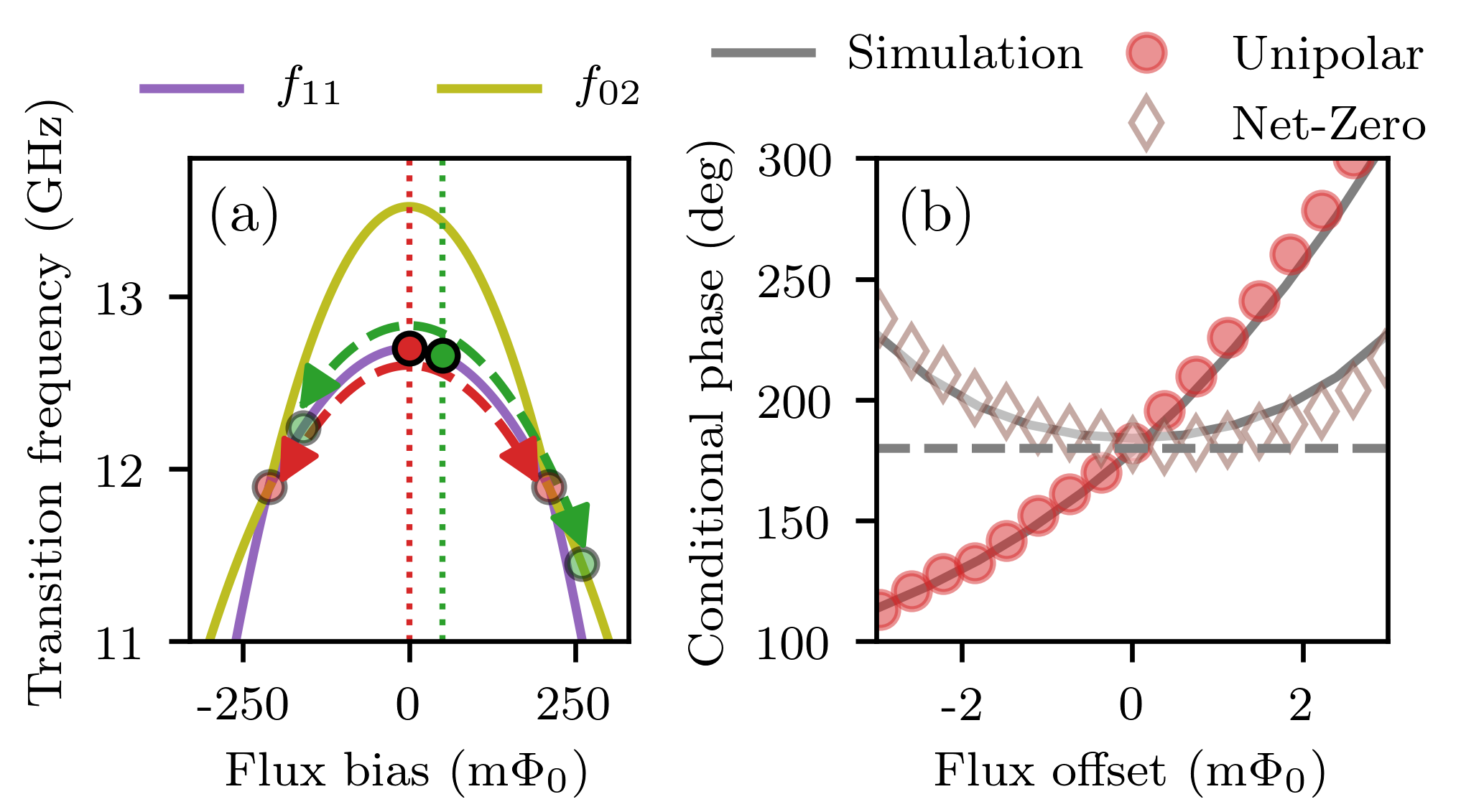}
    \caption{\label{fig:echo_effect}
    Echo effect in $\netzero$ pulses.
    (a) Level diagram showing the effect of a drift in flux on a $\netzero$ pulse:
    a $\netzero$ pulse will move to the interaction point on the positive and the negative arm (red); when the bias is offset (green), one arm will overshoot while the other arm will undershoot the interaction point, canceling the acquired extra phase to first order.
    (b) Measured dependence of conditional phase on applied DC flux offset for both $\netzero$ (diamond) and unipolar (circles) CZ pulses ($\TtwoQ=40~\ns$, $\ToneQ=20~\ns$).
    The grey lines correspond to simulation~\cite{Suppmaterial} and the dashed line indicates $180~\deg$.
    The unipolar ($\netzero$) is first-order (second-order) sensitive to the applied offset.
  }
\end{figure}

We next investigate a built-in echo effect that provides protection against flux noise.
Because the derivative of the flux arc is equal and opposite in sign at the positive and negative halves of the $\netzero$ pulse, we expect $\phi_{01}$ and $\condphase$ to be first-order insensitive to low-frequency flux noise.
As a test, we measure the dependence of  $\condphase$ on an applied DC flux offset for both a unipolar and $\netzero$ $\CZ$ gate~[\cref{fig:echo_effect}].
As shown in \cref{fig:echo_effect}(b), $\condphase$ is first-order sensitive for a unipolar pulse and only second-order sensitive for $\netzero$.
We have also measured how the dephasing time depends on the detuning for both a square flux pulse and two half-square flux pulses with opposite sign~\cite{Suppmaterial}.
We find that the dephasing rate is significantly reduced when the opposite-sign flux pulses are used, confirming that $\netzero$ pulses have a built-in echo effect.

The pulse shape is intended to minimize leakage and is described by two parameters~\cite{Suppmaterial}.
Parameter $\thetaf$ is a measure of the flux at the middle of the unipolar pulse, and at the middle of each half of $\netzero$.
States $\ket{11}$ and $\ket{02}$ are resonant at $\thetaf=\pi/2$.
Parameter $\lambda_2$ tunes the sharpness of the pulse rise and fall.
We follow~\cite{Wood18} in defining the leakage ($\leak$) of an operation as the average probability that a random computational state leaks out of the computational subspace.

\begin{figure}
  \centering
    \includegraphics{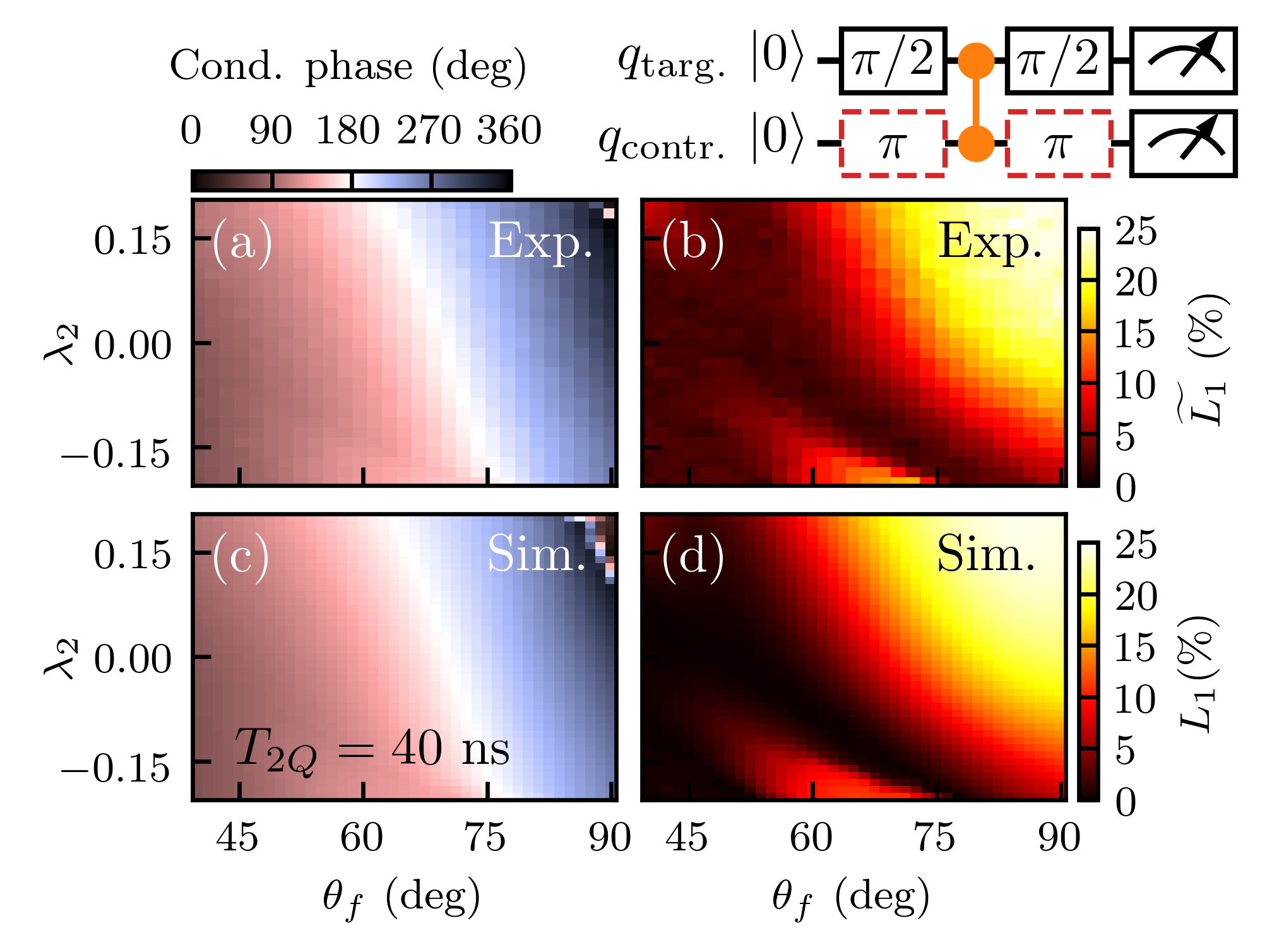}
    \caption{\label{fig:sim_comp}
    Conditional phase (a, c) and leakage (b, d) for a $\TtwoQ=40~\ns$ strong $\netzero$ flux pulse ($\ToneQ=20~\ns$) as a function of pulse parameters $\theta_f$ and $\lambda_2$ for both experiment (a, b) and simulation (c, d).
    The conditional phase increases with $\thetaf$ and $\lambda_2$, since both of these have the effect of making the pulse spend more time close to the interaction point.
    Leakage tends to increase significantly with larger values of $\thetaf$ with the exception of a diagonal fringe.
  }
\end{figure}

In order to gain insight into how $\condphase$ and $\leak$ depend on the pulse shape, we perform an experiment and compare this to simulations.
The conditional oscillation experiment~(\cref{fig:sim_comp}) consists of a Ramsey-like experiment that allows us to measure $\condphase$ and estimate $\leak$.
This experiment measures the phase acquired during an (uncalibrated) CZ gate by the target qubit ($\Qtarget$) while either leaving the control qubit ($\Qcontrol$) in the ground state, or adding and subsequently removing an excitation to $\Qcontrol$.
The difference between the phase acquired when $\Qcontrol$ is in $\ket{0}$ and when $\Qcontrol$ is in $\ket{1}$ gives $\condphase$.
If leakage from $\ket{11}$ to $\ket{02}$ occurs, $\Qcontrol$ is in $\ket{0}$ when the second $\pi$ pulse, intended to remove the excitation, is applied.
This will result in adding, instead of removing, an excitation to $\Qcontrol$.
The leakage probability $\leak$ can be estimated as $\estleak = m/2$, where $m$ is the population difference on the control qubit between both variants of the experiment.
Because of relaxation effects, $\estleak$ slightly overestimates~$\leak$.

The simulations model the system in a realistic way and allow us to extract $\condphase$, $\leak$ and the average gate fidelity $\fid$ for a single application of the gate~\cite{Suppmaterial}.
The pulse is modeled as a trajectory in a two-qutrit Hamiltonian.
The noise model accounts for relaxation and dephasing effects as well as the effect of the calibrated remaining distortions.
For the dephasing we take into account the different timescales on which flux noise acts as well as their dependence on the flux bias.
Distortions are modeled based on the remaining distortions measured using the Cryoscope technique~\cite{Rol19_cryoscope}.

Both the experiment and simulation show a fringe of low leakage~[\cref{fig:sim_comp}(b,d)].
This fringe can be understood by analogy to a Mach-Zehnder interferometer~\cite{Suppmaterial}.
The states $\ket{11}$ and $\ket{02}$ correspond to two paths of the interferometer.
The first part of the $\netzero$ pulse (red in \cref{fig:concept_net_zero}) corresponds to the first (imbalanced) beamsplitter of the interferometer.
In general, after the first beamsplitter most of the population remains in $\ket{11}$ but a small part is transferred to $\ket{02}$.
Pulsing through the sweetspot (green in \cref{fig:concept_net_zero}) corresponds to the arms of the interferometer.
The two paths are detuned by $\sim 800~\MHz$, causing a phase to be acquired before the paths are recombined at the second half of the $\netzero$ pulse (blue in \cref{fig:concept_net_zero}) corresponding to the second beamsplitter.
The phase difference between the two paths will cause interference that either enhances or suppresses the leakage to $\ket{02}$~\cite{Suppmaterial}.

Given the good correspondence between experiment and simulation~(\cref{fig:sim_comp}), we can use simulations to explore the parameter space  $(\thetaf, \lambda_2, \TtwoQ)$ to find the shortest $\TtwoQ$ and corresponding parameters enabling a high-fidelity, low-leakage $\CZ$ gate.
The minimum $\CZ$ gate duration is fundamentally limited by the coupling strength $J_2$ as the time required to acquire 180 degrees of conditional phase at the avoided crossing: $\TtwoQ\ge\frac{\pi}{J_2}=25~\ns$.
We find a $\TtwoQ = 28~\ns$ $\netzero$ pulse that makes use of the Mach-Zehnder interference condition to achieve a low leakage.
We append a $\ToneQ=12$~ns flux pulse to correct the single-qubit phases on both qubits, making the total duration of the phase-corrected $\CZ$ gate $\TCZ=40$~ns.
We ensure that these phase-correction pulses satisfy~\cref{eq:netzero_cond} and have a sufficiently low amplitude to not  affect $\condphase$ and $\leak$ significantly.

\begin{figure}
  \centering
    \includegraphics{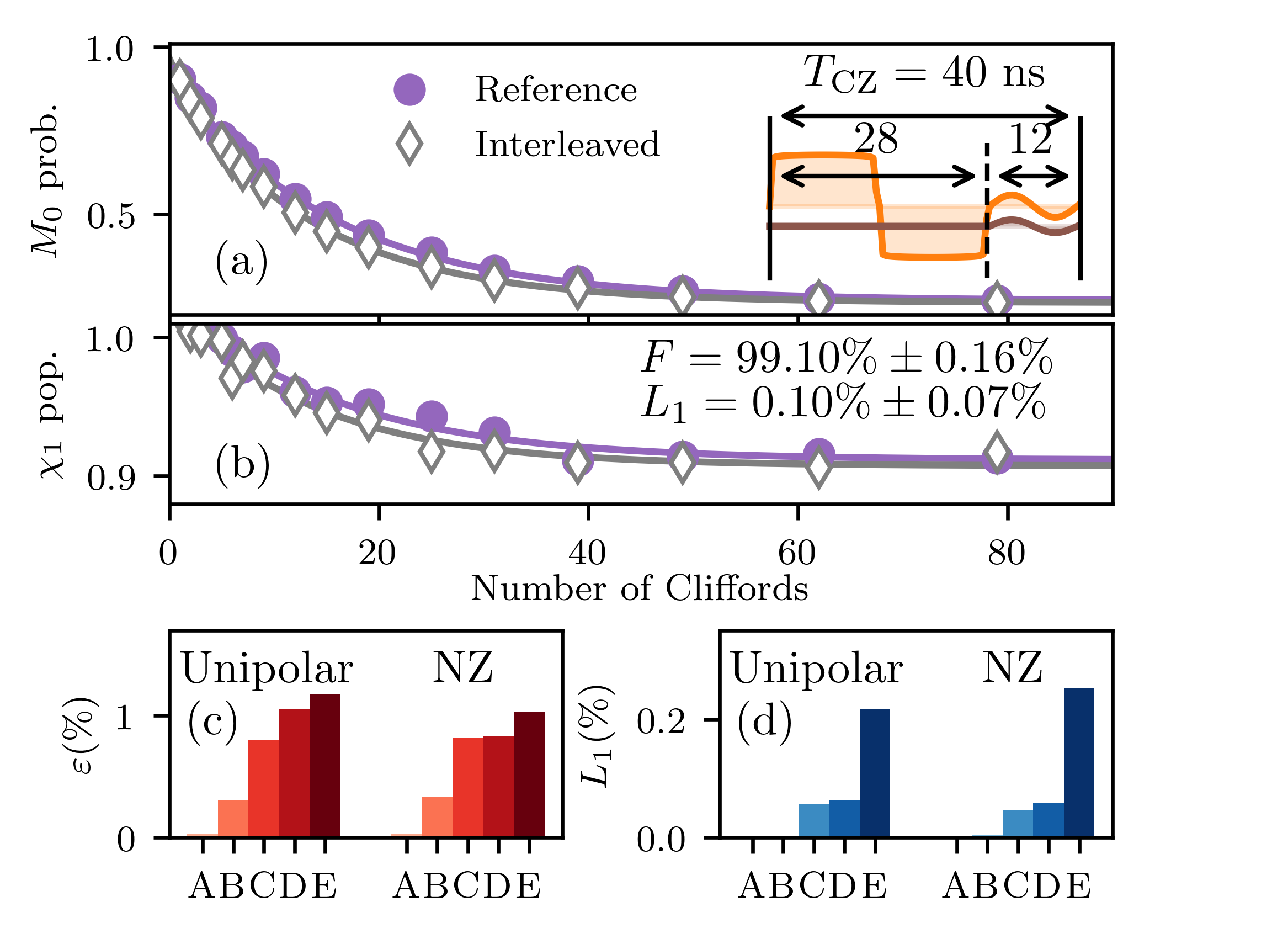}
    \caption{\label{fig:IRB} Interleaved randomized benchmarking with leakage modification and simulated performance using different error models for a $\TCZ=40~\mathrm{ns}$ $\CZ$ gate.
    (a) Survival probability $M_0$ of recovering $\ket{00}$ for the reference two-qubit randomized benchmarking sequence and the sequence that contains an interleaved $\netzero$ CZ gate between each Clifford.
    (b) Population in the computational subspace $\compsub$.
    The diagram schematically shows the pulses used to perform the $\CZ$ gate: a $\TtwoQ=28~\ns$ strong $\netzero$ pulse to acquire $\condphase$ followed by simultaneous weaker $\netzero$ pulses of duration $\ToneQ=12~\ns$ to correct $\phi_{01}$ and $\phi_{10}$.
    Simulated $\varepsilon$ (c) and $\leak$ (d) for different error models~\cite{Suppmaterial}.
    The error models ($A$ to $E$) contain: no noise ($A$), relaxation ($B$), all Markovian noise components ($C$), Markovian and quasi-static flux noise components ($D$) and all noise components including distortions ($E$).
  }
\end{figure}

We characterize the performance of the $\CZ$ gate using an interleaved randomized benchmarking protocol~\cite{Magesan12b, Barends14} with a few modifications that allow us to quantify leakage~\cite{Asaad16, Wood18, Suppmaterial}.
The randomized benchmarking sequences are based on 300 random seeds. For each seed, every data point is measured 104 times.
We measure an average gate fidelity $\fid=99.10\%\pm0.16\%$ and leakage $\leak=0.10\%\pm 0.07\%$ for the $\netzero$ pulse with $\TCZ=40~\ns$~[\cref{fig:IRB}(a,b)].
We did not perform similar measurements for the unipolar pulse as this gate is not repeatable.

It is possible to investigate the limits to the performance of the $\CZ$ gate using simulation.
We simulate this gate for a range of different error models~[\cref{fig:IRB}(c,d)].
A first observation is that the infidelity ($\infid=1-\fid$)~of the $\netzero$ gate does not significantly increase when the low-frequency flux-noise components are included in the simulations, whereas this does affect the unipolar pulse.
It appears that the difference in $\infid$ between the unipolar and $\netzero$ pulses for the full model including distortions can be attributed completely to this effect.
This observation is consistent with the echo effect demonstrated in \cref{fig:echo_effect}.
Looking at the $\leak$ error budgets, $\leak$ is limited by distortions.
This is understandable as minimizing $\leak$ requires the pulse to follow a precise trajectory.
The simulations also indicate that dephasing causes leakage.
This can be understood as dephasing effectively corresponds to an uncertainty in the energy of these levels.
The simulated $\leak$ is larger than the measured $\leak$.
This could be explained in two ways, either the distortions are less severe than our estimate, or the simulations, only concerned with a single application of the gate, do not take into account all the relevant effects.
Specifically, because the population in the leakage subspace does not completely decohere, this population can seep back into the computational subspace due to an interference effect (similar to that in the $\netzero$ pulse itself) at subsequent applications of the gate.
Because the first $\CZ$ gate cannot benefit from this coherence, the simulations, which only deal with a single $\CZ$ gate, slightly overestimate the effective leakage.

In summary, we have demonstrated a flux-based $\CZ$ gate for transmon qubits that is fast, low-leakage, high-fidelity and repeatable.
The gate is realized using a bipolar Net-Zero flux pulse that harnesses a Mach-Zehnder-like interference to achieve speed while reducing leakage by destructive interference.
The NZ pulse exploits the flux symmetry of the pulsed transmon to build in an echo effect on its single-qubit phase and the conditional phase, increasing fidelity relative to a unipolar pulse.
Finally, the action of the NZ pulse is robust to long-timescale distortions in the flux-control line remaining after real-time pre-compensation, enabling the repeatability of the $\CZ$ gate.
These features make the realized NZ CZ gate immediately useful in high-circuit-depth applications of a full-stack quantum computer in which a controller issues operations to execute on the quantum hardware in real time.
For example, current work in our group uses NZ CZ gates to stabilize two-qubit entanglement  by multi-round indirect parity measurements~\cite{Bultink19_ZZXX}.
Future work will incorporate NZ CZ gates into our scheme~\cite{Versluis17} to realize a surface-code-based logical qubit~\cite{Fowler12} with monolithic transmon-cQED quantum hardware.

\begin{acknowledgments}
We thank J.~Helsen, T.~O'Brien, X.~Bonet-Monroig, N.~Haandaek, Y.~Salathe,  and V.~Ostroukh for useful discussions.
This research is supported by the Office of the Director of National Intelligence (ODNI), Intelligence Advanced Research Projects Activity (IARPA),
via the U.S. Army Research Office grant W911NF-16-1-0071.
The views and conclusions contained herein are those of the authors and should not be interpreted as necessarily representing the official policies or endorsements, either expressed or implied, of the ODNI, IARPA, or the U.S. Government.
F.B. and B.M.T. are supported by ERC grant EQEC No. 682726.
\end{acknowledgments}

\clearpage
\onecolumngrid
\renewcommand{\theequation}{S\arabic{equation}}
\renewcommand{\thefigure}{S\arabic{figure}}
\renewcommand{\thetable}{S\arabic{table}}
\renewcommand{\bibnumfmt}[1]{[S#1]}
\renewcommand{\citenumfont}[1]{S#1}
\setcounter{figure}{0}
\setcounter{equation}{0}
\setcounter{table}{0}
\section*{Supplemental material for ``A fast, low-leakage, high-fidelity two-qubit gate for a programmable superconducting quantum computer''}
\date{\today}
\maketitle

This supplemental material contains detailed information on the experimental protocols and the simulations performed in this work.
\cref{sec:device_params} provides relevant device parameters.
\cref{sec:pulse_parametrization} describes the parametrization used for the unipolar and $\netzero$ pulses.
\cref{sec:simulation_structure} describes the simulations in detail.
\cref{sec:conditional_oscillation_exp} and \cref{sec:leakage_mod_RB} describe protocols used to characterize the flux pulses.
\cref{sec:optimal_performance} investigates the limitations of the $\CZ$ gate.
\cref{sec:interferometer} discusses the Mach-Zehnder interferometer analogy in detail.

\subsection{Device parameters}
\label{sec:device_params}
All experiments were performed on a circuit-QED quantum chip containing three starmon-type~\cite{Versluis17_SOM} transmon qubits, labeled $\QH$, $\QM$, and $\QL$. Pairs $\QH$-$\QM$ and $\QM$-$\QL$ are coupled by separate bus resonators.
Each qubit has a microwave drive line for single-qubit gating, a flux-bias line for local and ns-timescale control of the qubit frequency, and dedicated, fast readout resonators with Purcell protection for the qubits.
The readout resonators are coupled to a common feed line, allowing independent readout of the three qubits by frequency multiplexing.

In this work we focus on the transmon pair $\QH$-$\QM$.
We have achieved similar performance (fidelity, leakage and gate time) for the pair $\QM$-$\QL$.
Relevant device parameters are given in \cref{tab:device_parameters}.

\begin{table}[!h]
   \begin{tabular}{|c|cccccc|}
      \hline
      \multicolumn{1}{c}{} &~~~~  &~~~~ &~~~~  &~~~~ &~~~~ &~~~~ \\[\dimexpr-\normalbaselineskip-\arrayrulewidth]
      \textbf{Parameter} & \multicolumn{2}{c|}{$\QL$}& \multicolumn{2}{c|}{$\QM$}& \multicolumn{2}{c|}{$\QH$}\\
      \hline \hline
      {$\omega/2\pi$ operating point (GHz)} & \multicolumn{2}{c|}{5.02}& \multicolumn{2}{c|}{5.79}& \multicolumn{2}{c|}{6.87}\\
      \hline
      {$\omega/2\pi$ sweetspot (GHz)} & \multicolumn{2}{c|}{5.02}& \multicolumn{2}{c|}{5.79}& \multicolumn{2}{c|}{6.91}\\
      \hline
      {$\anharmonicity/2\pi$ (MHz)} & \multicolumn{2}{c|}{-300}& \multicolumn{2}{c|}{-300}& \multicolumn{2}{c|}{-331}\\
      \hline
      {$J_1/2\pi$ avoided crossing (MHz)} & \multicolumn{3}{c|}{17.2}& \multicolumn{3}{c|}{14.3}\\
      \hline
      {$\Tone$ $(\us)$} & \multicolumn{2}{c|}{31.8}& \multicolumn{2}{c|}{15.2}& \multicolumn{2}{c|}{19.2}\\
      \hline
      {$\Ttwos$ operating point $(\us)$} & \multicolumn{2}{c|}{14.0}& \multicolumn{2}{c|}{14.8}& \multicolumn{2}{c|}{3.2}\\
      \hline
      {$T_2^{\echo}$ operating point $(\us)$} & \multicolumn{2}{c|}{33.8}& \multicolumn{2}{c|}{19.4}& \multicolumn{2}{c|}{14.7}\\
      \hline
      {$\sim\omega_\mathrm{bus}/2\pi$ (GHz)} & \multicolumn{3}{c|}{8.5}& \multicolumn{3}{c|}{8.5}\\
      \hline
    \end{tabular}
    \caption{
        \label{tab:device_parameters}
        Parameters of the three-transmon device: qubit frequency ($\omega$), anharmonicity ($\anharmonicity$), exchange coupling between $\ket{01}$ and $\ket{10}$ ($J_1$), dephasing times ($T_1, T_2^\ramsey, T_2^{\echo}$) and bus-resonator frequency ($\omega_\mathrm{bus}$).
        Experiments in this work are performed with the pair $\QH$-$\QM$.
        $\QH$ is operated $40~\MHz$ below its sweetspot to minimize interaction with a spurious two-level system right at the sweetspot frequency.
}
\end{table}

\subsection{Flux pulse parametrization}
\label{sec:pulse_parametrization}
Unipolar and $\netzero$ pulses are based on the Martinis-Geller parametrization for fast-adiabatic gates~\cite{Martinis14_SOM}.
This parametrization is determined by the Hamiltonian [\cref{eq:hamiltonian}] projected onto a two-dimensional subspace.
In the case of the $\CZ$~gate, this subspace is spanned by the states $\ket{11}$ and  $\ket{02}$.
The projected Hamiltonian, $\qubitHamiltonian$, takes the form
\begin{equation}
\qubitHamiltonian = \begin{pmatrix} \frac{\detuning}{2} & J_2\\ J_2&-\frac{\detuning}{2} \end{pmatrix},
\end{equation}
where $\detuning=\omega_{\ket{02}}-\omega_{\ket{11}}$ is the bare detuning between $\ket{11}$ and $\ket{02}$ and $J_2$ is their coupling.
The detuning $\detuning$ is controlled by flux whereas $J_2$ is considered to be constant.
We define the angle $\theta$ as
\begin{equation}
\label{eq:angle}
\theta \equiv \arctan \left(\frac{2J_2}{\detuning}\right).
\end{equation}
Note that $\theta=\pi/2$ at $\detuning = 0$.

The waveform is expressed as a series
\begin{equation}
\label{eq:martinis_geller}
\theta(\tau(t)) = \theta_i + \sum_{j=1}^N \lambda_j \left(1-\cos\left(\frac{2\pi\cdot j\cdot \tau(t)}{\TtwoQ}\right)\right),
\end{equation}
where $\TtwoQ$ is the pulse duration, $\theta_i$ corresponds to the detuning at the operating point and $\tau$ is proper time, which is related to real time $t$ through $t(\tau) = \int^\tau_0 d\tau' \sin \left(\theta(\tau ')\right)$.

We truncate the series to $N=2$. We make use of the relation between the angle at the middle of the unipolar pulse ($\thetaf$) and the odd $\lambda$ coefficients
\begin{equation}
\thetaf\equiv\theta(\TtwoQ/2)=\theta_i+2\sum_{j \:\text{odd}}^N \lambda_j,
\end{equation}
to define the entire waveform using three parameters: $\thetaf, \lambda_2$, and $\TtwoQ$. A $\netzero$ pulse is a sequence of two concatenated unipolar pulses, each lasting $\TtwoQ/2$ time and with the same $\thetaf$ and $\lambda_2$.

\begin{figure}
  \centering
    \includegraphics{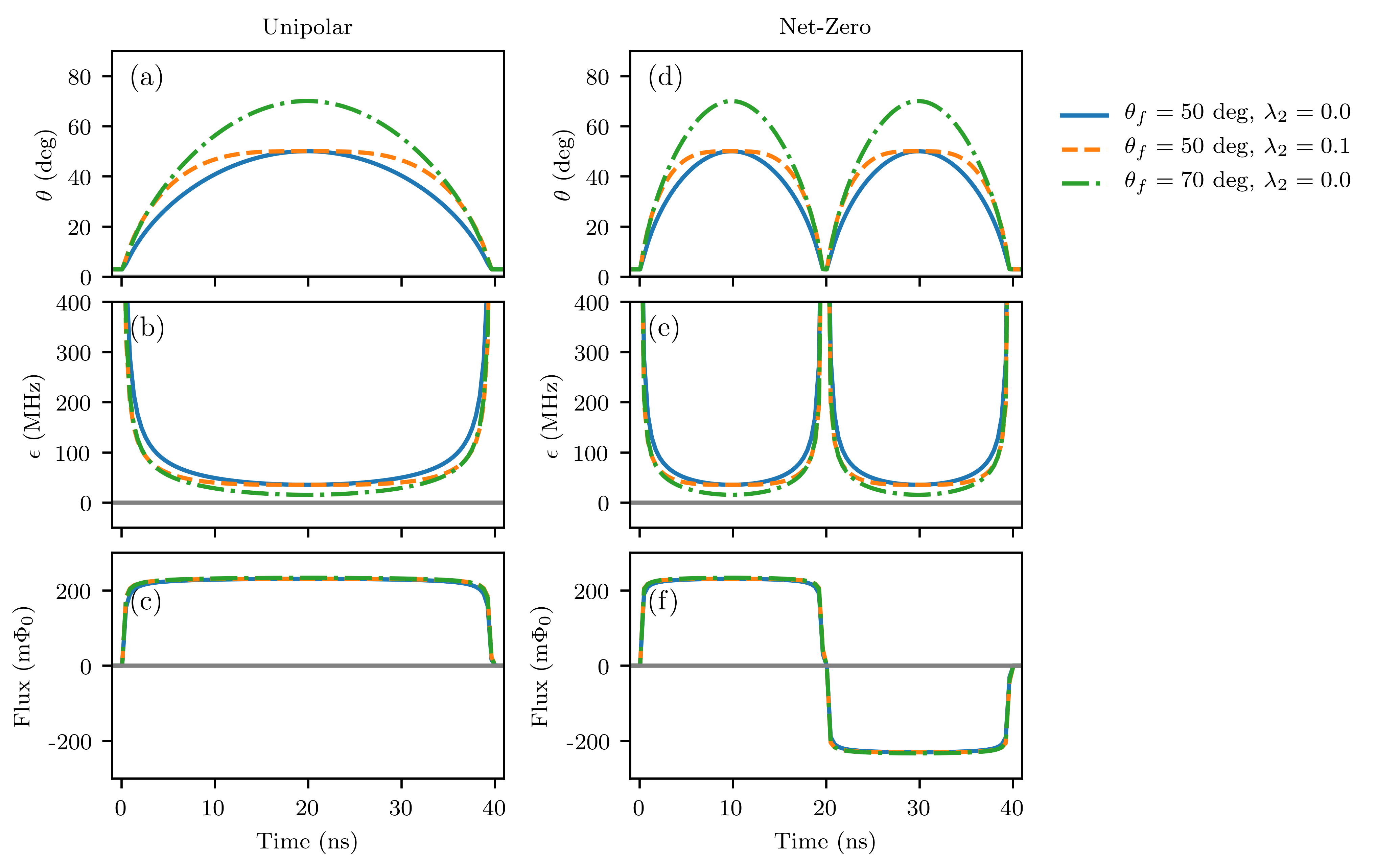}
    \caption{\label{fig:pulse_parametrization}
    Unipolar (a-c) and $\netzero$ pulses (d-e) represented in terms of $\theta$ (a, d), bare detuning $\detuning$ (b, e) and flux $\Phi$ (c, f).
    The center of the unipolar pulse is controlled by $\thetaf$, while $\lambda_2$ controls the sharpness of rise and fall of the pulse.
  }
\end{figure}

There are a few more transformations required in order to have a waveform in terms of the flux $\targetFlux(t)$ [\cref{fig:pulse_parametrization}]:
\begin{equation}
\theta(t) \mapsto \detuning(t) \mapsto \omega_{\QH}(t) \mapsto \targetFlux(t).
\end{equation}
The first transformation uses \cref{eq:angle}: $\epsilon(t)=2J_2/\tan\theta(t)$.
The second one uses the fact that by definition $\detuning(t)=\omega_{\ket{02}}(t)-\omega_{\ket{11}}(t)=\omega_{\QH}(t)+\anharmonicity_{\QH}-\omega_{\QM}$.
The qubit frequency depends on flux according to the formula
\begin{equation}
    \omega_{\QH}(\Phi) = (\omega_{\QH}^0-\anharmonicity_{\QH})\sqrt{\left|\cos\Bigl( \frac{\Phi}{\Phi_0}\pi\Bigr)\right|} + \anharmonicity_{\QH},
    \label{eq:flux_arc}
\end{equation}
where $\omega_{\QH}^0$ is the sweetspot frequency and $\anharmonicity_{\QH}$ the anharmonicity, reported in~\cref{tab:device_parameters}.
We refer to this relation between frequency and flux as the flux arc. The flux arc has been measured in the experiment and we find that it matches well with~\cref{eq:flux_arc}.
We invert~\cref{eq:flux_arc} to convert $\omega_{\QH}(t) \mapsto \targetFlux(t)$. Since $\omega_{\QH}(\Phi)=\omega_{\QH}(-\Phi)$, there is a positive and a negative solution for every value of $\omega_{\QH}$. In the case of a unipolar pulse, we always consider the positive solution, whereas, in the case of a $\netzero$ pulse, the first and second half of the pulse use the positive and negative solutions, respectively.
Changes that are clearly visible in the $\theta$ parametrization correspond to only a small change in the applied flux.
This provides intuition why even a small distortion of the applied flux can have a relatively large effect on the gate quality.

\subsection{Simulation structure}
\label{sec:simulation_structure}
\begin{figure}
    \centering
    \includegraphics{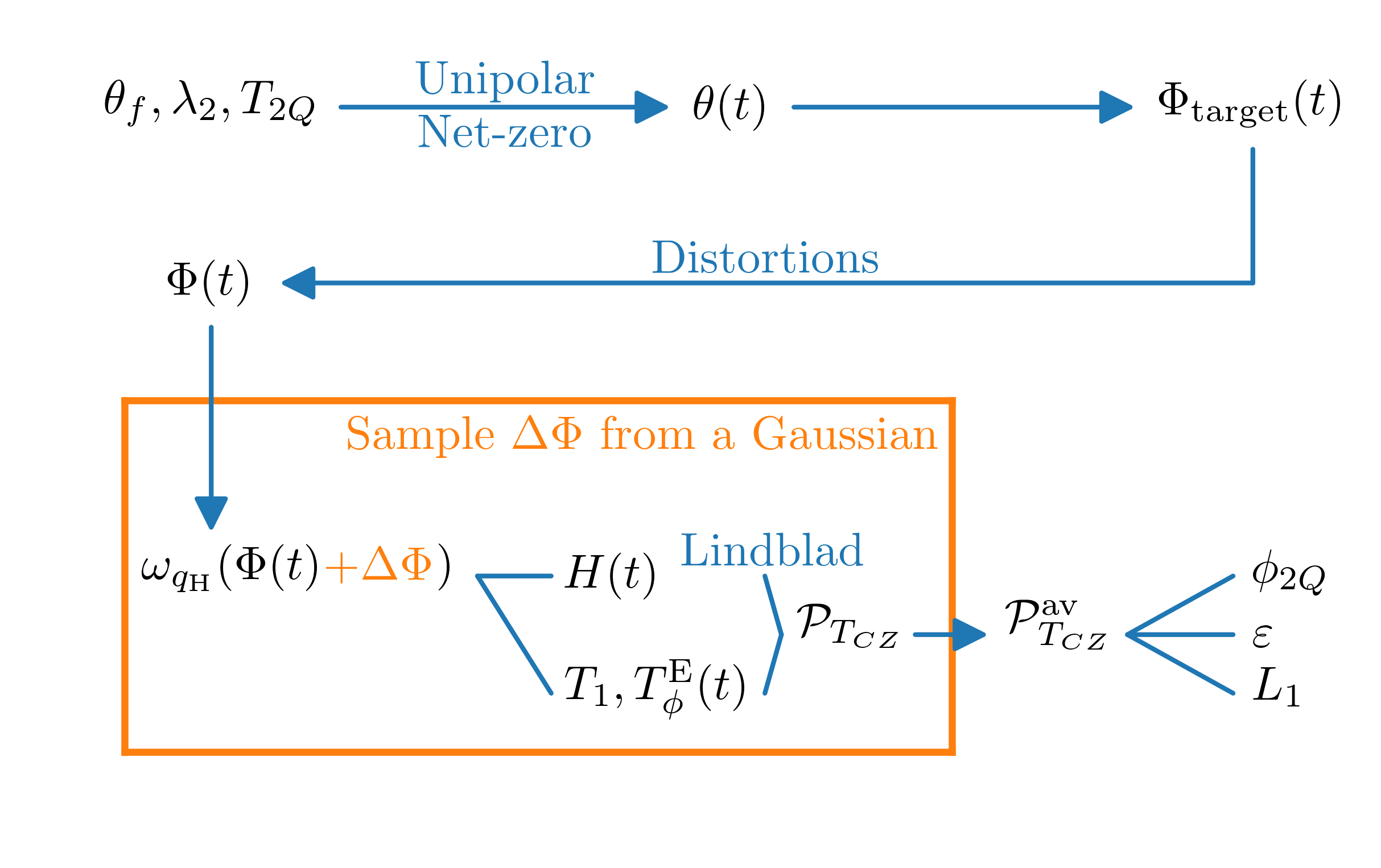}
    \caption{\label{fig:general_simulation_structure}
        The parameters $\thetaf,\lambda_2$ and the gate time $\TtwoQ$ determine either a unipolar pulse or a $\netzero$ pulse in terms of $\theta(t)$, see \cref{eq:martinis_geller}.
        $\theta(t)$ is converted into $\PhiT(t)$ thorough various transformations described in~\cref{sec:pulse_parametrization}.
        Pulse distortions are applied by convolution to compute $\Phi(t)$ experienced by the qubit.
        The solution of the Lindblad equation is the time-evolution superoperator $\mathcal{P}_{\TCZ}$. Averaging over a Gaussian distribution for the quasi-static flux bias $\Delta\Phi$, we obtain the average superoperator $\mathcal{P}_{\TCZ}^\mathrm{av}$. From that any quantity of interest can be computed, in particular the conditional phase $\condphase$, the average gate infidelity $\infid$ and the leakage $\leak$.
    }
\end{figure}

The simulations model the system, consisting of two coupled transmons, using a two-qutrit Hamiltonian.
One of the two transmons, namely $\QH$, is actively pulsed into resonance according to the pulse parametrization described in \cref{sec:pulse_parametrization}.
The simulations~(\cref{fig:general_simulation_structure}) include distortions, relaxation and flux-dependent dephasing effects.
The error model also includes a distinction between Markovian (fast) and non-Markovian (slow) noise in order to accurately model dephasing effects.
The simulations are used to calculate the {propagator} or time-evolution superoperator, from which the quantities of interest - fidelity, leakage and conditional phase - are extracted.

\subsubsection{System Hamiltonian}

The system is composed of two transmons coupled via a bus resonator. We exclude the resonator from the model by making the assumption that it always remains in its ground state (it is excited only ``virtually'').
We restrict each transmon to its first three energy levels.
In the dispersive regime, in the rotating-wave approximation, the Hamiltonian is given by
\begin{align}
H(t) =\:\, &\omega_{\LSQ} a_{\LSQ}^\dagger a_{\LSQ} + \frac{\anharmonicity_{\LSQ}}{2} (a_{\LSQ}^\dagger)^2 a_{\LSQ}^2 +
\omega_{\MSQ}(\actualFlux(t))\, a_{\MSQ}^\dagger a_{\MSQ} + \frac{\anharmonicity_{\MSQ}}{2} (a_{\MSQ}^\dagger)^2 a_{\MSQ}^2 \\
&+ J_1(\actualFlux(t))\, (a_{\LSQ} a_{\MSQ}^\dagger + a_{\LSQ}^\dagger a_{\MSQ}),
\label{eq:hamiltonian}
\end{align}
where only the higher-frequency transmon ($\MSQ$) is actively fluxed. Here $a_{\LorMSQ}$ is the annihilation operator restricted to the first three energy levels, $\omega_{\LorMSQ}$ and $\anharmonicity_{\LorMSQ}$ are the qubit frequency and anharmonicity respectively, and $J_1$ is the coupling.
The coupling is weakly flux-dependent since $J_1(\Phi)\approx \frac{g_{\LSQ}g_{\MSQ}}{2}(\Delta_{\LSQ}^{-1}+\Delta_{\MSQ}^{-1}(\Phi))$, with $g_{\LorMSQ}$ the coupling of $\LorMSQ$ to the bus resonator and $\Delta_{\LorMSQ}\approx\omega_\text{bus}-\omega_{\LorMSQ}\gg g_{\LorMSQ}$ given the parameters in~\cref{tab:device_parameters}.
When we generate the flux pulse according to \Cref{sec:pulse_parametrization}, we consider $J_2=\sqrt{2}J_1$ to be constant and $J_2$ equal to its measured value at the $\ket{11} \leftrightarrow \ket{02}$ avoided crossing, whereas in the simulations we take into account the dependence of $J_1$ and $J_2$ on $\Phi$.

\subsubsection{Distortions}
\begin{figure}
    \centering
    \includegraphics{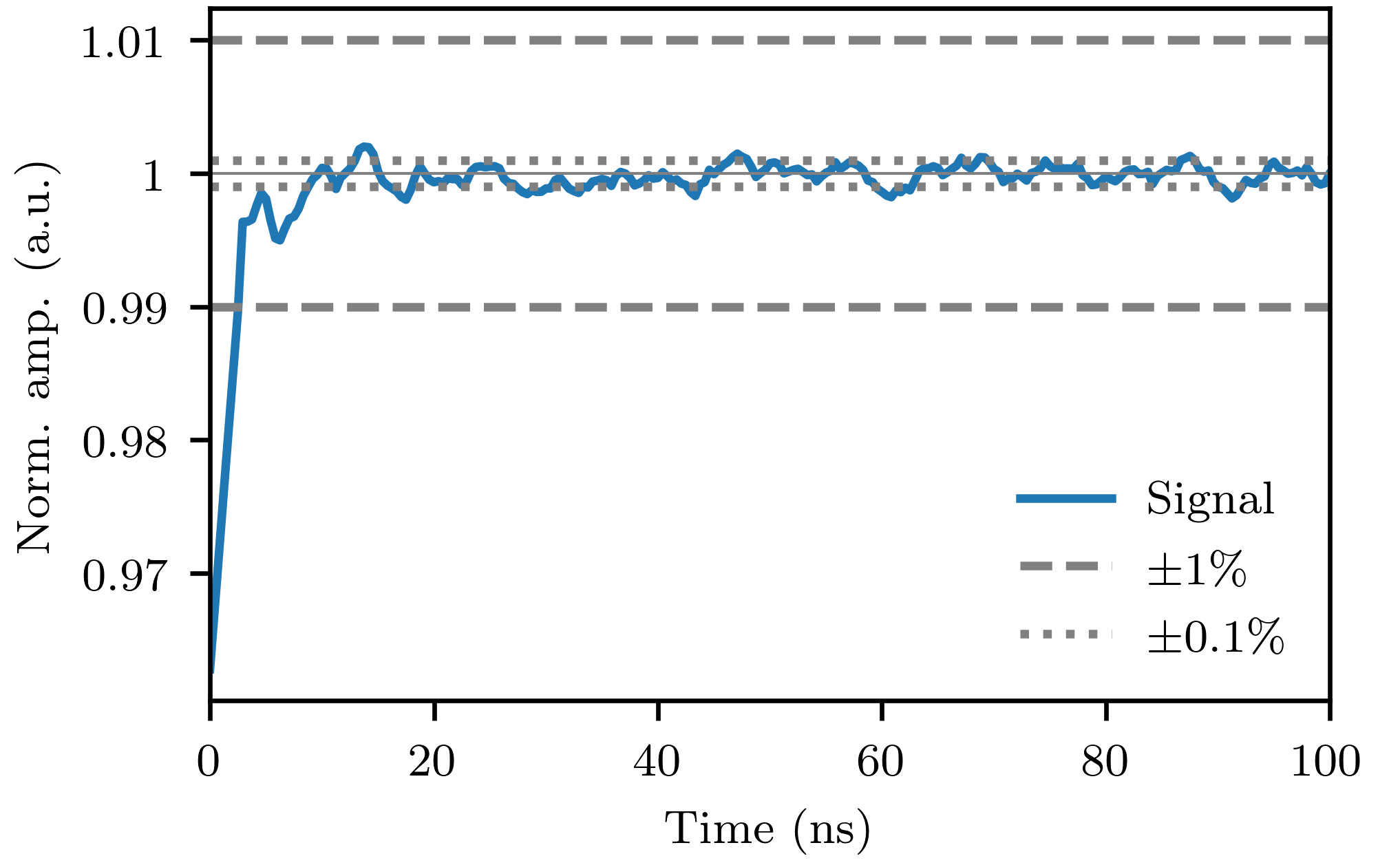}
    \caption{\label{fig:cryoscope}
        Step response at the qubit after applying distortion corrections, measured using the Cryoscope technique~\cite{Rol19_cryoscope_SOM}.
        The impulse response extracted from this experiment is used to distort the pulses in the simulations.
        In the case of perfect distortion corrections, the normalized amplitude would have value 1 for all times larger than zero.
    }
\end{figure}

The flux pulse at the qubit is subject to distortions altering the shape of the waveform as experienced by the qubit.
Distortions are described as a linear time-invariant system fully characterized by the impulse response $\impulseresponse$ of the system.
We best compensate such distortions by predistorting the desired pulse $\targetFlux(t)$ with an impulse response $\tilde{\impulseresponse}^{-1}$ designed to invert $\impulseresponse$.
Then, the actual pulse $\actualFlux(t)$ experienced by the qubit is given by
\begin{equation}
\actualFlux(t) = (\impulseresponse \ast \VAWG )(t) = (\impulseresponse \ast (\tilde{\impulseresponse}^{-1} \ast \PhiT) )(t) =((\tilde{\impulseresponse}^{-1} \ast \impulseresponse) \ast \PhiT )(t),
\end{equation}
where $\ast$ denotes convolution.
The distortions remaining after applying $\tilde{\impulseresponse}^{-1}$ are determined by measuring the step response $s(t) = \int_{0}^{t}dt'\,(\tilde{\impulseresponse}^{-1} \ast \impulseresponse)(t')$ (\cref{fig:cryoscope}) using the Cryoscope technique~\cite{Rol19_cryoscope_SOM}.
The impulse response extracted from this data is used to distort the pulses in simulations.

\subsubsection{Noise model}

There are two major error sources in superconducting qubits: relaxation and flux noise.
The latter has a power spectral density $S_f\sim A/f$, where $f$ is frequency and $\sqrt{A}$ is a constant of the order of 10 $\mu\Phi_0$, with $\Phi_0$ the flux quantum.
$S_f$ contains both high-frequency and low-frequency components: we phenomenologically distinguish high and low frequencies depending on whether they are larger or smaller than $1/\TCZ$. Relaxation and high-frequency flux-noise components are Markovian noise processes since they act on a timescale shorter than the gate time. On the other hand, the low-frequency flux-noise components determine a non-Markovian noise process, since they induce correlations across different gates.

\begin{figure}
    \centering
    \includegraphics{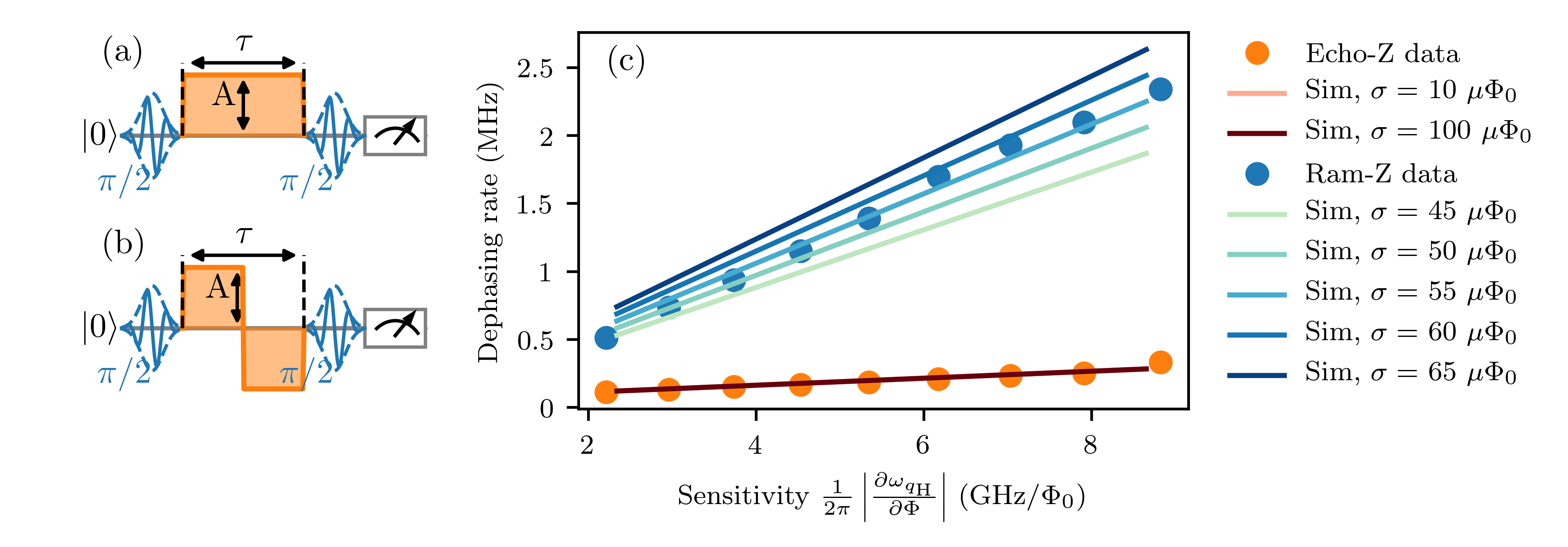}
    \caption{\label{fig:matching_sigma}
        Comparison of experimental data and simulation (c) for the Ram-Z (a) and Echo-Z (b) experiments.
        In the Ram-Z (Echo-Z) experiment, the dephasing time is measured using a (two-half) square flux-pulse(s).
        All simulated curves include the effects of both the Markovian and non-Markovian noise.
        Only the strength of the non-Markovian noise~[\cref{eq:average_propagator}], quantified by $\sigma$, is varied, while the strength of the Markovian noise, quantified by $T_{1,\QH}$ and $T_{\phi,\QH}^{\echo}(\Phi)$, is kept fixed.
        We see that the value $\sigma=55\:\mu\Phi_0$ best fits the Ram-Z data. It fits the Echo-Z data as well, given that the simulated curves are equal even for $\sigma$'s that differ by an order of magnitude. This agrees with the intuition that the non-Markovian noise is echoed-out in an Echo-Z experiment.
    }
\end{figure}

We perform two experiments to quantify the strength of the dephasing affecting $\QH$: a Ram-Z and an Echo-Z experiment~[\cref{fig:matching_sigma}].
In these experiments, the dephasing times $T_{2,\MSQ}^{\ramsey}(\Phi)$ and $T_{2,\MSQ}^{\echo}(\Phi)$, respectively, at different flux sensitivities $\frac{1}{2\pi}\frac{\partial \omega_{\MSQ}}{\partial \Phi}$ are measured while applying a flux pulse.
In the Ram-Z experiment, this flux pulse is square.
In the Echo-Z experiment, the flux pulse consists of two square half pulses that detune the qubit by the same amount in magnitude but with opposite-sign sensitivity.
We perform these experiments for a range of fluxes.
The experimental data for $\QH$ is represented in~\cref{fig:matching_sigma}.
On the other hand, the static qubit $\QM$ is always operated at the sweetspot.
Therefore, we only use the measured Ramsey and Echo dephasing times at the sweetspot~\cref{tab:device_parameters}.
The relaxation times $T_{1,\QH}$ and $T_{1,\QM}$, are also reported in~\cref{tab:device_parameters}.

We assume that the low-frequency flux-noise components are echoed out in an Echo-Z experiment. In other words, we assume that $T_{1,\LorMSQ}$, $T_{2,\LorMSQ}^{\echo}(\Phi)$ quantify the strength of the Markovian noise. On the other hand, we assume that $T_{1,\LorMSQ}$, $T_{2,\LorMSQ}^{\ramsey}(\Phi)$ quantify the strength of the overall noise (both Markovian and non-Markovian). The strength of the non-Markovian noise alone cannot be extracted directly from the experiment. However, in the following we explain the model that we use fitting the experimental data~(\cref{fig:matching_sigma}). In this way we can simulate separately both the Markovian and non-Markovian noise, and obtain a realistic simulation of the system.

\emph{Model of Markovian noise.}\\
A Markovian evolution is modeled with the Lindblad equation
\begin{gather}
\dot{\rho}(t) = -i [H(t),\rho(t)] + \sum_{j,\LorMSQ} \Bigl(c_{j,\LorMSQ}(t)\rho(t)c_{j,\LorMSQ}^\dagger(t)-\frac{1}{2} \{c_{j,\LorMSQ}^\dagger(t)c_{j,\LorMSQ}(t),\rho(t)\} \Bigr) \eqqcolon \mathcal{L}_t\bigl(\rho(t)\bigr), \label{eq:Lindbladequ}
\end{gather}
where $\mathcal{L}_t$ is the time-dependent Lindbladian defined by the Hamiltonian~[\cref{eq:hamiltonian}] and by the jump operators $\{c_{j,\LorMSQ}(t)\}$ specified in~\cref{eq:collapse_op_relaxation,eq:collapse_op_dephasing_1,eq:collapse_op_dephasing_2,eq:collapse_op_dephasing_3} below.

To model relaxation, we use the jump operator
\begin{equation}
c_{0,\LorMSQ} = \sqrt{\frac{1}{{T}_{1,\LorMSQ}}} a_{\LorMSQ}.
\label{eq:collapse_op_relaxation}
\end{equation}
To model pure dephasing, we first define a pure-dephasing time
\begin{equation}
T_{\phi,\LorMSQ}^{\echo}(\Phi) = \left( \frac{1}{T_{2,\LorMSQ}^{\echo}(\Phi)} - \frac{1}{{2T}_{1,\LorMSQ}} \right)^{-1},
\end{equation}

Ignoring relaxation-induced dephasing in this paragraph, the coherence $\braket{0}{\rho_{\LorMSQ}(\Phi)|1}$ decays as $e^{-t/T_{\phi,\LorMSQ}^{\echo}(\Phi)}$, where $\rho_{\LorMSQ}$ is the qutrit reduced density matrix.
In~\cref{fig:matching_sigma} we see that the decay rates have a linear dependence on the flux sensitivity.
Ignoring the anharmonicity, the frequency of the $\ket{2}$ state is twice the frequency of the $\ket{1}$ state, therefore, the sensitivity of the $\ket{2}$ state is twice as high.
Given these two observations, we assume that $\braket{0}{\rho_{\LorMSQ}(\Phi)|2}\propto e^{-t/(T_{\phi,\LorMSQ}^{\echo}(\Phi)/2)}$ and $\braket{1}{\rho_{\LorMSQ}(\Phi)|2} \propto e^{-t/T_{\phi,\LorMSQ}^{\echo}(\Phi)}$.
We find that such decay rates can be realized by the following jump operators

\begin{align}
c_{1,\LorMSQ}(\Phi(t)) &= \sqrt{\frac{8}{9{T}_{\phi,\LorMSQ}^{\echo}(\Phi(t))}} \begin{pmatrix}
1 & 0 & 0 \\
0 & 0 & 0 \\
0 & 0 & -1
\end{pmatrix}_{\LorMSQ},  \label{eq:collapse_op_dephasing_1} \\
c_{2,\LorMSQ}(\Phi(t)) &= \sqrt{\frac{2}{9{T}_{\phi,\LorMSQ}^{\echo}(\Phi(t))}} \begin{pmatrix}
1 & 0 & 0 \\
0 & -1 & 0 \\
0 & 0 & 0
\end{pmatrix}_{\LorMSQ},  \label{eq:collapse_op_dephasing_2} \\
c_{3,\LorMSQ}(\Phi(t)) &= \sqrt{\frac{2}{9{T}_{\phi,\LorMSQ}^{\echo}(\Phi(t))}} \begin{pmatrix}
0 & 0 & 0 \\
0 & 1 & 0 \\
0 & 0 & -1
\end{pmatrix}_{\LorMSQ}. \label{eq:collapse_op_dephasing_3}
\end{align}

Instead, if one would use only
\begin{equation}
c'_{1,\LorMSQ}(\Phi(t)) = \sqrt{\frac{2}{{T}_{\phi,\LorMSQ}^{\echo}(\Phi(t))}} \begin{pmatrix}
1 & 0 & 0 \\
0 & 0 & 0 \\
0 & 0 & -1
\end{pmatrix}_{\LorMSQ},   \label{eq:std_dephasing_model}
\end{equation}
which produces the same Lindbladian as $c''_{1,\LorMSQ}(\Phi(t)) = \sqrt{{2}/{T}_{\phi,\LorMSQ}^{\echo}(\Phi(t))} \, a_{\LorMSQ}^\dagger a_{\LorMSQ}$,
then one would get $\braket{0}{\rho_{\LorMSQ}(\Phi)|2}\propto e^{-t/(T_{\phi,\LorMSQ}^{\echo}(\Phi)/4)}$ and $\braket{1}{\rho_{\LorMSQ}(\Phi)|2} \propto e^{-t/T_{\phi,\LorMSQ}^{\echo}(\Phi)}$. This means that~\cref{eq:std_dephasing_model} would be the correct modeling if the decay rates in~\cref{fig:matching_sigma} would depend quadratically on the sensitivity, but they do not.

The formal solution of~\cref{eq:Lindbladequ} is given by
\begin{equation}
\rho(t)=\mathcal{T} e^{\int_0^t dt'\,\mathcal{L}_{t'}} \:\bigl(\rho(0)\bigr),
\end{equation}
where $\mathcal{T}$ is the time-ordering operator.
We call $\mathcal{P}_{\TCZ}\coloneqq\mathcal{T} e^{\int_0^{\TtwoQ} dt'\,\mathcal{L}_{t'}}$ the {propagator} or time-evolution superoperator, evaluated up to the gate time ${\TCZ}$, which includes an idling time $\ToneQ$ to account for the noise during the single-qubit phase correction pulses.
The propagator $\mathcal{P}_{\TCZ}$ can be computed by solving the differential~\cref{eq:Lindbladequ}, or as
\begin{equation}
\mathcal{P}_{\TCZ} \simeq e^{\delta t \mathcal{L}_{{\TCZ}-\delta t}}
\, e^{\delta t \mathcal{L}_{{\TCZ}-2\delta t}} \ldots\: e^{\delta t \mathcal{L}_{2\delta t}} \, e^{\delta t \mathcal{L}_{\delta t}} \, e^{\delta t \mathcal{L}_{0}}, \label{eq:discretetimeevolution}
\end{equation}
for a sufficiently small $\delta t$. In the simulations we use $\delta t = 0.1$ ns.
In the Liouville representation, this equation is a product of matrices.
We find that this method is an order of magnitude faster than using the qutip~\cite{Johansson13} differential equation solver.

\emph{Model of non-Markovian noise.}\\
We model the low-frequency flux-noise components as quasi-static.
Since the static qubit~$\QM$ is always operated at the sweetspot, where the sensitivity to flux noise is zero, we apply this model only to~$\QH$.
We assume that the qubit experiences a random, fixed flux offset $\Delta\Phi$ during the execution of a gate, but that $\Delta\Phi$ varies across different gates.
For $\Delta\Phi\ll 1$, the effect of such offset on the pulse trajectory can be approximated at first order as $\omega_{\MSQ}(\Phi(t)+\Delta\Phi)\approx \omega_{\MSQ}(\Phi(t))+\frac{\partial \omega_{\MSQ}(\Phi(t))}{\partial \Phi} \Delta\Phi$, where $\frac{1}{2\pi}\frac{\partial \omega_{\MSQ}(\Phi)}{\partial \Phi}$ is the flux sensitivity.
Using~\cref{eq:flux_arc} we can see that $\frac{\partial \omega_{\MSQ}(\Phi)}{\partial \Phi}=-\frac{\partial \omega_{\MSQ}(-\Phi)}{\partial \Phi}$.
In the case of a $\netzero$ pulse, this implies that first-order frequency variations in the first half of the pulse are canceled by an equal and opposite variation in the second half, resulting in an echo effect.

We take the probability distribution $p_{\sigma}$ of $\Delta\Phi$ to be Gaussian $p_{\sigma}(\Delta\Phi)=e^{-(\Delta\Phi)^2/(2\sigma^2)}/(\sqrt{2\pi}\sigma)$, where $\sigma$ is the standard deviation of the Gaussian.
Averaging over this distribution, we get the final propagator
\begin{equation}
\mathcal{P}_{\TCZ}^\mathrm{av}=\int_{-\infty}^{+\infty}d(\Delta\Phi)\:\: p_{\sigma}(\Delta\Phi)\cdot \mathcal{P}_{\TCZ}(\Delta\Phi), \label{eq:average_propagator}
\end{equation}
which gives the time evolution including all the noise sources in the model, both Markovian and non-Markovian.

The standard deviation $\sigma$ is not directly measured in the experiment. Instead, we fit this model to the experiment simulating a Ram-Z and Echo-Z experiment for~$\QH$~(\cref{fig:matching_sigma}). We vary the value of $\sigma$ while keeping the Markovian noise model described above fixed. We find that the value $\sigma=55~\mu\Phi_0$ best fits both the Ram-Z and Echo-Z data at the same time. This is the value we use in all the simulations in this paper.

\subsubsection{Quantities of interest}

To quantify the quality of the $\CZ$ gate, we are interested in computing the conditional phase, the leakage and the average gate fidelity from the propagator $\mathcal{P}_{\TCZ}^\mathrm{av}$.
In the following, we summarize their definitions for a generic superoperator $\mathcal{P}$.

We call $\compsub$ the computational subspace, spanned by the 2-qubit energy levels $\ket{00},\ket{01},\ket{10}$ and $\ket{11}$ at the operating point. The phases acquired by those states under the action of $\mathcal{P}$ are computed as
\begin{equation}
e^{i\phi_{ij}}=\frac{\bra{ij}\mathcal{P}\bigl( \ket{ij}\bra{00} \bigr)\ket{00}}{\abs{\bra{ij}\mathcal{P}\bigl( \ket{ij}\bra{00} \bigr)\ket{00}}}, \label{eq:qoi_phases}
\end{equation}
where $i,j\in\{0,1\}$.
If $\mathcal{P}$ is unitary, that is, $\mathcal{P}(\rho)=U\rho U^\dagger$ for some unitary $U$, then~\cref{eq:qoi_phases} reduces to $e^{i\phi_{ij}}=\frac{\braket{ij}{U|ij}}{\abs{\braket{ij}{U|ij}}}$, and, if $U$ is diagonal, then we simply have $U\ket{ij}=e^{i\phi_{ij}}\ket{ij}$.
The phase $\phi_{00}$ of the ground state can be set to 0.
The single-qubit phases are given by $\phi_{01}$ and $\phi_{10}$.
The  conditional phase $\condphase$ is defined as the phase acquired by the target qubit conditional on the state of the control qubit and it is given by
\begin{equation}
\condphase = \phi_{11} - \phi_{10} - \phi_{01}.
\end{equation}
Note that  $\condphase$ is invariant under single-qubit $Z$ rotations.

We follow the definitions in~\cite{Wood18_SOM} for leakage, seepage and average gate fidelity.
The leakage of a superoperator~$\mathcal{P}$ is defined as
\begin{align}
\leak &= 1 - \int_{\psi_1\in\compsub} d\psi_1\: \tr_{\compsub}\Bigl(\mathcal{P}\bigl(\ket{\psi_1}\bra{\psi_1}\bigr)\Bigr) \label{eq:leakage}\\
&= 1- \frac{1}{\dim\compsub} \sum_{i,j\in\{0,1\}} \tr_{\compsub}\Bigl(\mathcal{P}\bigl(\ket{ij}\bra{ij}\bigr)\Bigr). \nonumber
\end{align}
The quantity $\leak$ represents the average probability that a random computational state leaks out of $\compsub$.

The seepage of a superoperator~$\mathcal{P}$  is defined as
\begin{align}
\seep &= 1 - \int_{\psi_2\in\leaksub} d\psi_2\: \tr_{\leaksub}\Bigl(\mathcal{P}\bigl(\ket{\psi_2}\bra{\psi_2}\bigr)\Bigr), \label{eq:seepage}
\end{align}
where $\leaksub$ is the leakage subspace.

The average gate fidelity, evaluated in the computational subspace, between $\mathcal{P}$ and a target unitary $U$ is defined as
\begin{align}
\fid &= \int_{\psi_1\in\compsub} d\psi_1\, \bra{\psi_1} U^\dagger \mathcal{P}\bigl( \ket{\psi_1}\bra{\psi_1} \bigr) U \ket{\psi_1} \\
&=\frac{\dim{\compsub}(1-\leak)+\sum_k \abs{\tr_{\compsub}(U^\dagger A_k)}^2}{\dim{\compsub}(\dim{\compsub}+1)}, \nonumber
\label{eq:avgatefid_compsub}
\end{align}
where the $\{A_k\}$ are the Kraus operators of $\mathcal{P}$.
The average gate infidelity is $\infid=1-\fid$.

\subsection{Conditional oscillation experiment}
\label{sec:conditional_oscillation_exp}

\begin{figure}
  \centering
    \includegraphics{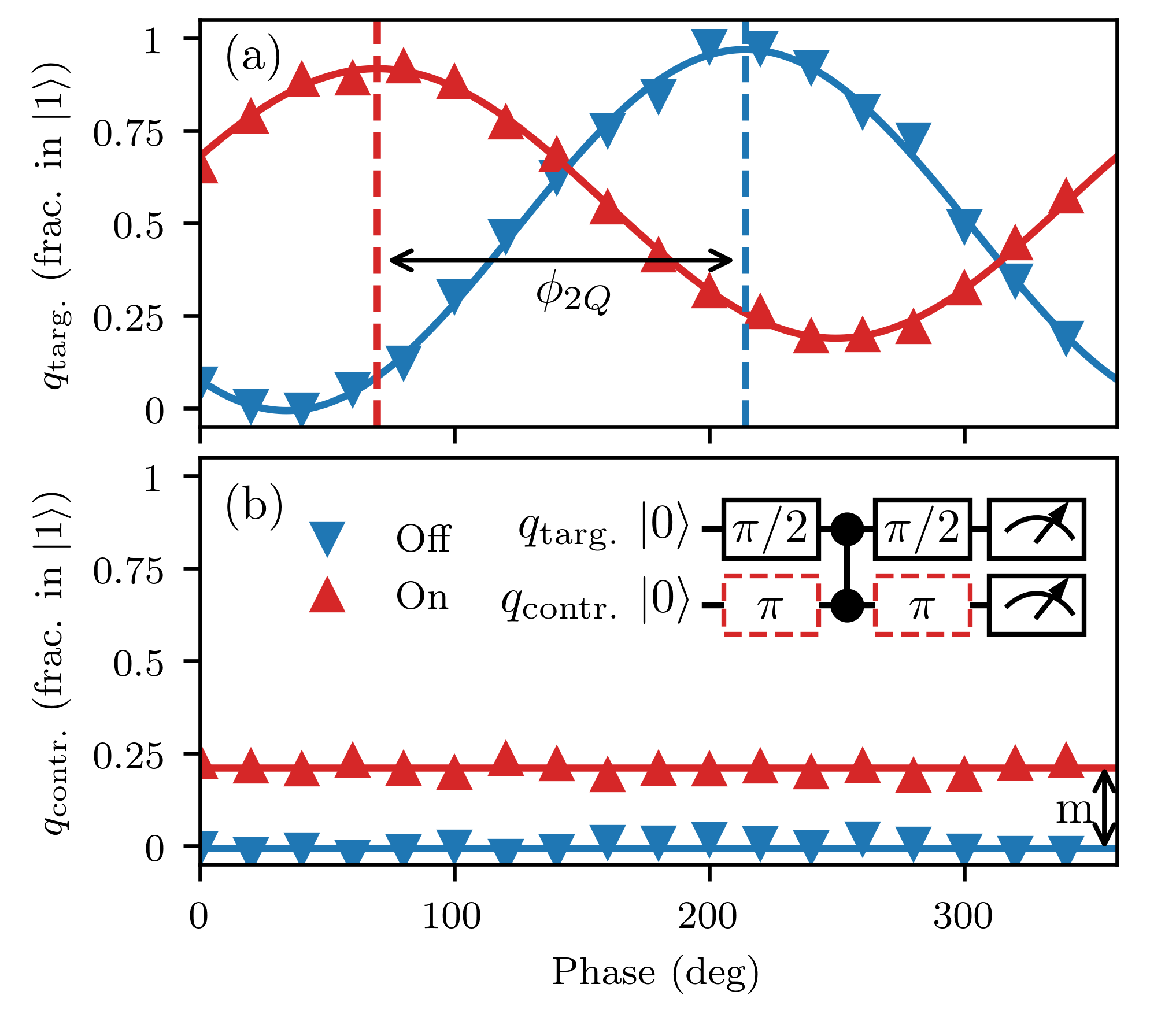}
    \caption{\label{fig:conditional_oscillation}
    The conditional oscillation experiment described in~\cref{sec:conditional_oscillation_exp}.
  }
\end{figure}

The conditional oscillation experiment~(\cref{fig:conditional_oscillation}) can be used to measure the single-qubit phases ($\phi_{01}$ and $\phi_{10}$) and the conditional phase ($\condphase$), and to estimate the leakage ($\leak$) defined in~\cref{eq:leakage}.
In the conditional oscillation experiment, two variants of the same experiment are performed.
In the first variant (Off), the target qubit ($\Qtarget$) is rotated onto the equator of the Bloch sphere by a $\pi / 2$ pulse and the control qubit ($\Qcontrol$) is left in the ground state.
After that, a flux pulse is applied that is intended to perform a $\CZ$ gate.
A recovery $\pi/2$ rotation, performed around an axis in the equatorial plane forming an angle $\phi$ with the $X$ axis, is applied to $\Qtarget$ before measuring the state of both qubits simultaneously.
In the second variant (On), $\Qcontrol$ is rotated into the excited state before applying the $\CZ$ gate.
Then, $\Qcontrol$ is pulsed back to the ground state before measuring both qubits.

The conditional phase $\condphase$ can be extracted directly from the phase of the oscillations and corresponds to the difference in phase between the oscillations~(\Cref{fig:conditional_oscillation}).
The single-qubit phase $\phi_{10}$ ($\phi_{01}$) can be measured by letting $\LSQ$ ($\MSQ$) take the role of $\Qtarget$ and correspond directly to the measured phase of $\Qtarget$ in the Off variant.

The quantity denoted by $m$ in \Cref{fig:conditional_oscillation} is called the missing fraction.
In the idealized case in which there is no noise and no leakage to other levels, we calculate $\leak=m_\mathrm{idealized}/2$. We see numerically that such relation approximately holds in the complete modeling with noise. Therefore, we define a leakage estimator $\estleak = \missingfrac/2$, where $m$ is the measured value. Due to relaxation effects, $\estleak$ generally overestimates $\leak$.
The advantage of estimating the leakage with $\estleak$ rather than with a randomized benchmarking experiment~(\cref{sec:leakage_mod_RB}) is that it is much faster. In this way we can quickly acquire a scan of the leakage landscape to find pulse parameters giving a low-leakage $\CZ$ gate. Further characterization is then carried out with randomized benchmarking.

\subsection{Optimal performance}
\label{sec:optimal_performance}
\begin{figure}
    \centering
    \includegraphics{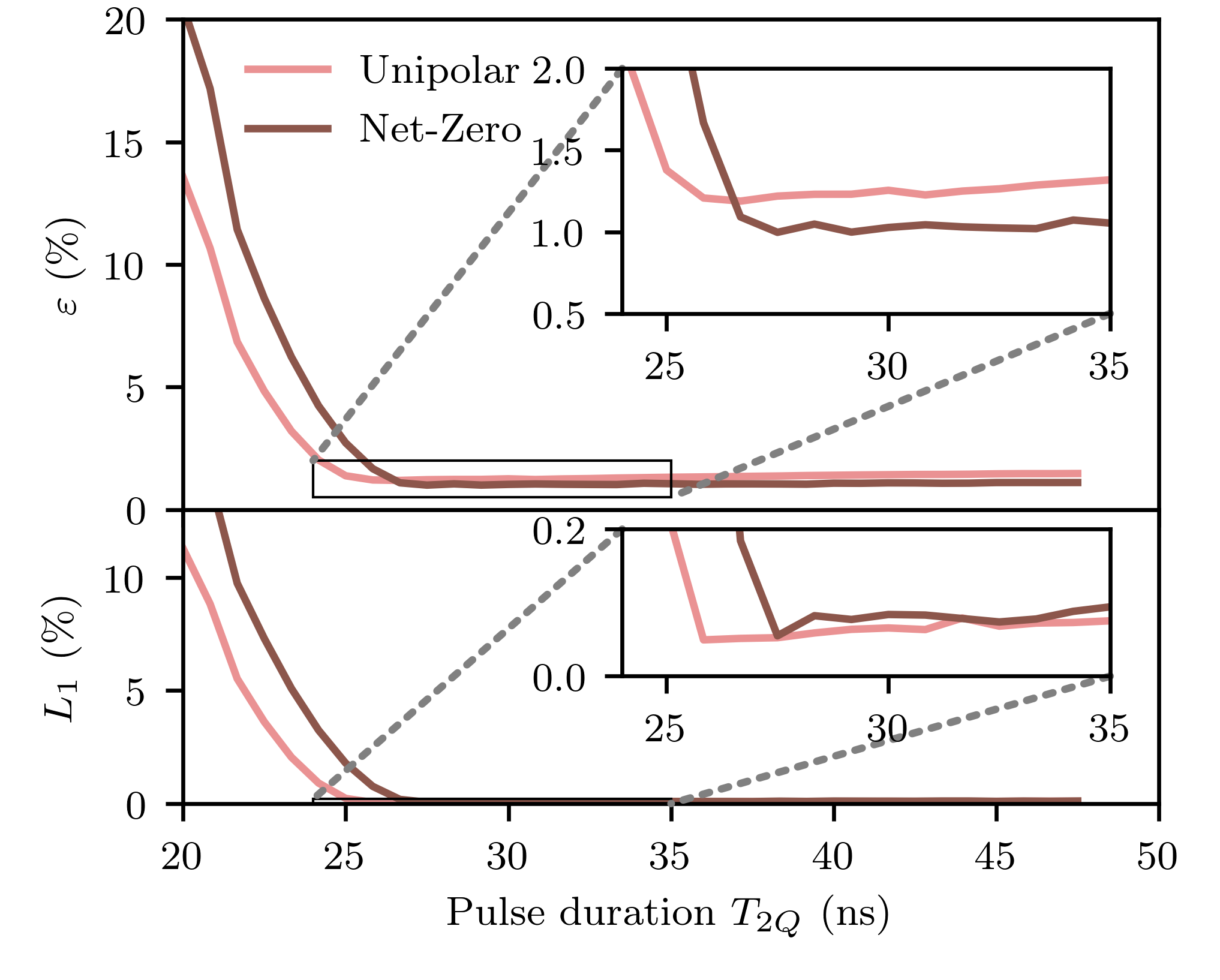}
    \caption{\label{fig:performance_pulse_duration}
    Minimal infidelity ($\infid$), optimized over $\thetaf$ and $\lambda_2$ for a fixed $\TtwoQ$, and leakage ($\leak$) evaluated at the minimal infidelity. Contrary to all other figures in this paper, the simulations shown here do not include distortions because we want to quantify the intrinsic optimal performance of unipolar and $\netzero$ pulses against Markovian and non-Markovian noise.
    We see that both $\infid$ and $\leak$ decrease fast approaching the speed limit $\pi/J_2\sim 25~\ns$ . Then $\netzero$ achieves lower infidelity and we can attribute this to the echo effect. We can use these simulations to find that the minimal $\TtwoQ$ to realize a high-fidelity, low-leakage $\netzero$ pulse is $\TtwoQ=28~\ns$.
    }
\end{figure}

Using simulations, it is possible to find the optimal parameters ($\thetaf$ and $\lambda_2$) for a given $\TtwoQ$ in order to perform a $\CZ$ gate.
We optimize over the infidelity $\infid$.
In \cref{fig:performance_pulse_duration}, the minimal infidelity $\infid$ and the corresponding leakage $\leak$ are shown as a function of $\TtwoQ$.
Contrary to all the other figures in this paper, the simulations shown in \cref{fig:performance_pulse_duration} do not include the effect of distortions.
The shortest duration for which a $\netzero$ pulse with low leakage and high fidelity can be performed is $\TtwoQ=28~\ns$, close to the speed limit of $\TtwoQ=25~\ns$, set by the interaction strength.
The difference in minimal infidelity between the unipolar and the $\netzero$ pulse is attributed to the built-in echo effect that makes the $\netzero$ pulse resilient to low-frequency flux-noise components.
Unipolar pulses with good performance could in principle be realized slightly faster ($\TtwoQ=26~\ns$) than $\netzero$ pulses, due to the fact that $\netzero$ needs $\sim 2~\ns$ to sweep from one avoided crossing to the other in the middle of the pulse, during which no conditional phase is accumulated.
However, unipolar pulses are not repeatable in practice.

\begin{figure}
    \centering
    \includegraphics{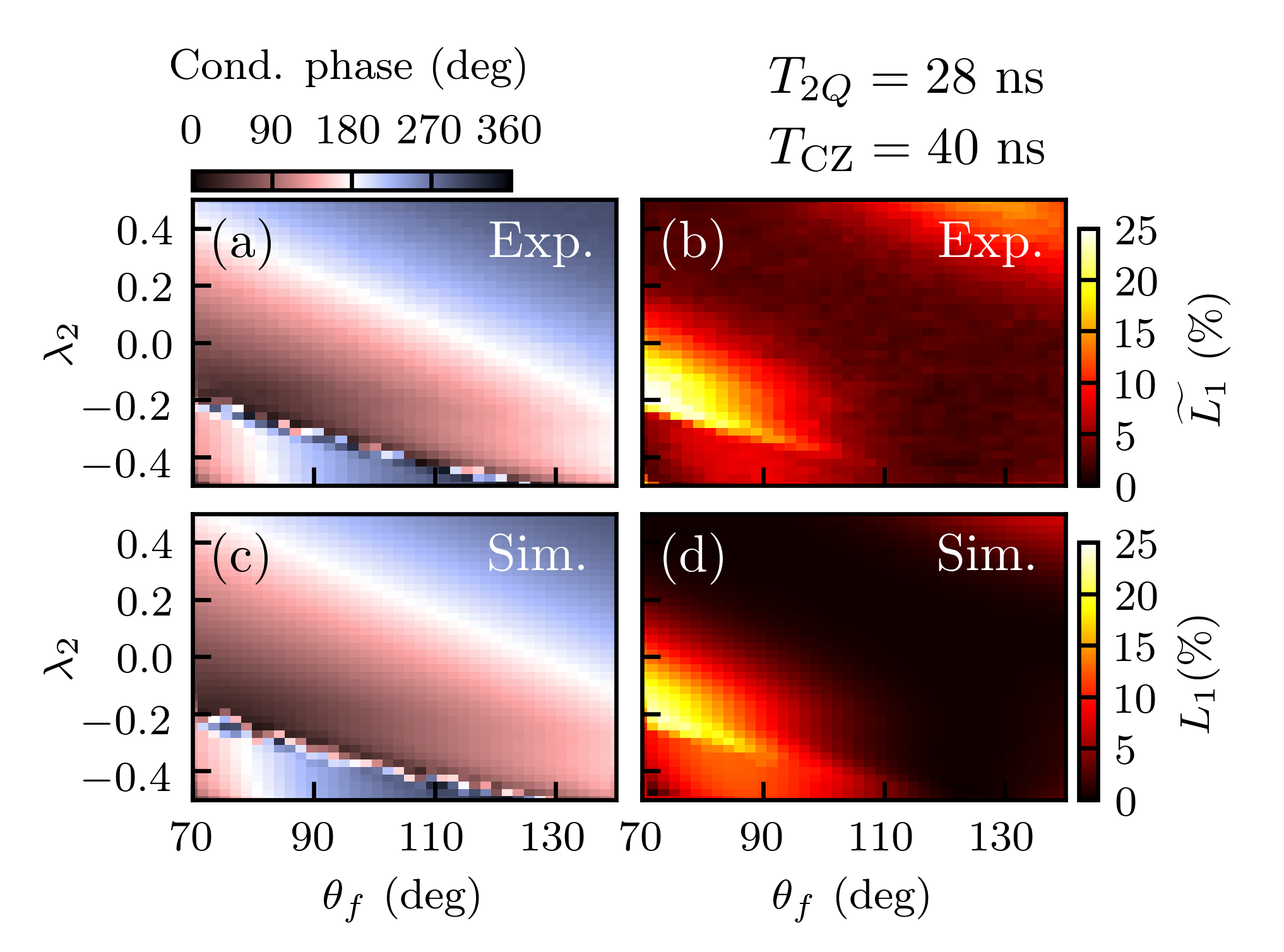}
    \caption{\label{fig:28ns_match}
    Matching of experimental (a,b) and simulated (c,d) landscapes of conditional phase and leakage as a function of the parameters $\thetaf$ and $\lambda_2$ of a $\TtwoQ=28~\ns$ strong $\netzero$ pulse ($\ToneQ=12~\ns$).
    The matching is excellent and in both cases we find the same broad region of low leakage.
    }
\end{figure}

The simulated landscape of the shortest duration ($\TtwoQ=28~\ns$) high-fidelity low-leakage $\netzero$ pulse is compared to experiment in~\cref{fig:28ns_match}.
There is a relatively large region of low-leakage at high $\thetaf$ (90-130~deg) that can be found in both simulation and experiment.
The $\TtwoQ=28~\ns$ pulses described in the main text are operating in this condition.

\subsection{Net-Zero pulses as a Mach-Zehnder interferometer}
\label{sec:interferometer}
To better understand the working of a $\netzero$ pulse, it is helpful to draw an analogy to a Mach-Zehnder interferometer.
In a $\netzero$ pulse, the trajectory first approaches the $\ket{11}\leftrightarrow\ket{02}$ avoided crossing at positive flux amplitude, then it sweeps through the sweetspot, and it finally goes in and out of the $\ket{11}\leftrightarrow\ket{02}$ avoided crossing at negative flux amplitude.
We argue that those three parts of the pulse correspond respectively to an (unbalanced) beamsplitter, to the arms of an interferometer, and to another (identical) beamsplitter.
We make a few idealizations in this analysis. Namely, we ignore the weak coupling to other states and we consider a purely unitary process.
Moreover, there is not a clear-cut separation between the beamsplitters, where the qubits are strongly coupled, and the arms of the interferometer, where they are effectively uncoupled.
However, since the sweep in the middle is very fast, for the sake of this model it does not really matter where the line is drawn.

In general, a unipolar pulse has the following effect on the $\ket{11}$ state
\begin{equation}
    \ket{11} \mapsto e^{i\BSphaseHalf}\sqrt{1-\alpha^2} \ket{11} + \alpha \ket{02},
\end{equation}
where $\alpha\in\mathbb{R}$ and $\alpha^2=4\leakHalf$ (assuming no leakage to other states).
In other words, during the first half of a $\netzero$ pulse, $\ket{11}$ acquires a certain conditional phase $\BSphaseHalf$  and it can also leak to $\ket{02}$, for example if the parameters of the pulse are not properly chosen or if the pulse is too short.

Unitarity implies that $\ket{02} \mapsto \alpha \ket{11} - e^{-i\BSphaseHalf}\sqrt{1-\alpha^2} \ket{02}$.
Overall, modulo a global phase, this amounts to the unitary
\begin{equation}
    B_1 = \begin{pmatrix}
        e^{i\BSphaseHalf}\sqrt{1-\alpha^2}  &  \alpha   \\
        \alpha     &    - e^{-i\BSphaseHalf}\sqrt{1-\alpha^2}
    \end{pmatrix},
\end{equation}
which is a beamsplitter that also imparts a conditional phase.

During the sweep across the sweetspot, $\ket{11}$ and $\ket{02}$ quickly acquire a relative phase $\PSphase$ due to the large energy gap between them ($\sim800$ MHz).
We can formalize this with the unitary
\begin{equation}
P_\PSphase = \begin{pmatrix}
1  &  0   \\
0     &    e^{i\PSphase}
\end{pmatrix},
\end{equation}
which is a phase shifter.

The second beamsplitter, $B_2$, is equal to $B_1$ due to the symmetry of the pulse. The total evolution is given by
\begin{equation}
B_2P_\PSphase B_1 = B_1P_\PSphase B_1 = \begin{pmatrix}
e^{i2\BSphaseHalf} \Bigl( (1-\alpha^2)+\alpha^2 e^{i\tilde{\PSphase}} \Bigr)  &  \alpha\sqrt{1-\alpha^2}e^{i\BSphaseHalf}(1-e^{i\tilde{\PSphase}})   \\
\alpha\sqrt{1-\alpha^2}e^{i\BSphaseHalf}(1-e^{i\tilde{\PSphase}})     &    \alpha^2 + (1-\alpha^2)e^{i\tilde{\PSphase}}
\end{pmatrix},
\end{equation}
where $\tilde{\PSphase}\coloneqq \PSphase-2\BSphaseHalf$.
We are interested in the first matrix element because it gives the leakage $\leakNZ$ and conditional phase $\BSphaseNZ$ at the end of a $\netzero$ pulse. Explicitly
\begin{gather}
    \leakNZ = \bigl( \alpha^4 + (1-\alpha^2)^2 + 2\alpha^2(1-\alpha^2) \cos\tilde{\PSphase}  \bigr)/4, \\
    \BSphaseNZ = 2\BSphaseHalf + \arctan \Biggl( \frac{\alpha^2 \sin\tilde{\PSphase}}{(1-\alpha^2) + \alpha^2\cos\tilde{\PSphase}} \Biggr). \label{eq:phicond_NZ_interferometer}.
\end{gather}
There are two cases in which $\leakNZ$ can be made zero. The first one is when $\alpha^2=0$. This is when the half pulse has zero leakage in the first place. We refer to this case as the adiabatic condition. The second case is when $\alpha^2\neq0$ but $\tilde{\PSphase}=(2k+1)\pi$, with $k$ an integer. We refer to this second case as the interference condition. We point out that, in either case, the second term in~\cref{eq:phicond_NZ_interferometer} is zero, which implies that $\BSphaseNZ = 2\BSphaseHalf$ whenever $\leakNZ=0$. As a consequence, the speed limit to do a $\netzero$ $\CZ$ with low leakage is the same as for the unipolar pulse ($\pi/J_2$).

\begin{figure}
    \centering
    \includegraphics{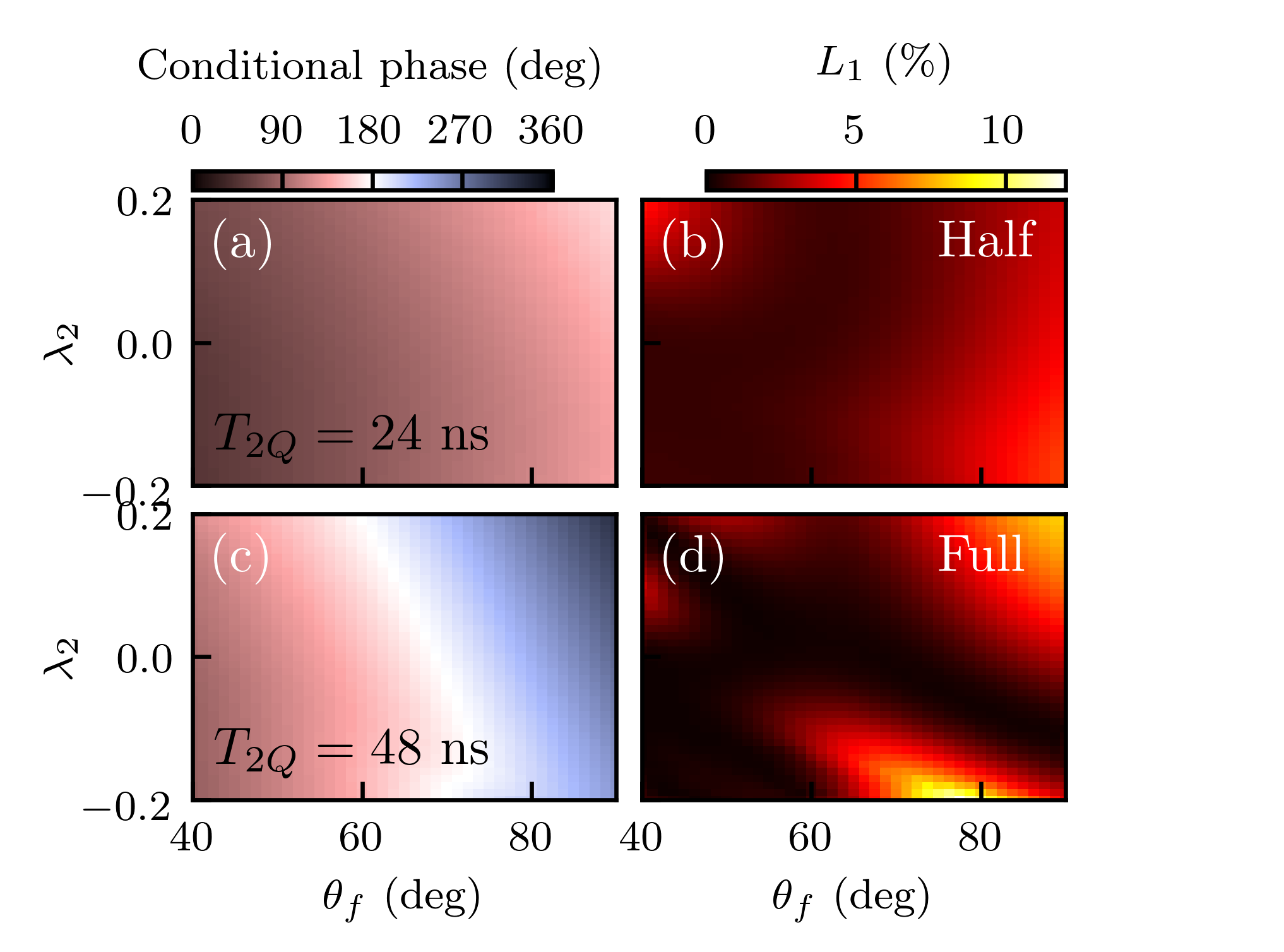}
    \caption{\label{fig:crosshair_sim}
    Simulation of conditional phase and leakage landscapes as a function of the parameters $\thetaf$, $\lambda_2$ of a half (a,b) and full (c,d) strong  $\netzero$ pulse $\TtwoQ=48~\ns$ ($\TCZ=60~\ns$).
    The half pulse consists of only the first part of the $\netzero$ pulse, which is effectively a $\TtwoQ=24~\ns$ unipolar pulse.
    Naively one may expect both the conditional phase and the leakage of the full pulse to be approximately twice that of the half-pulse.
    However, this is not the case for the leakage.
    In~(b) we see a low-leakage area due to the adiabaticity of the pulse.
    We find this low-leakage area in~(d) as well.
    However, an interference fringe is visible that does not occur for the half pulse.
    }
\end{figure}

It is possible to explore both the adiabatic and interference conditions for low leakage in the simulations~(\cref{fig:crosshair_sim}).
When performing a $\TtwoQ=24~\ns$ unipolar (Half $\netzero$) pulse, only the adiabatic condition can be used to achieve a low leakage.
This condition is visible as the dark region in~[\cref{fig:crosshair_sim}(b)].
When simulating a $\TtwoQ=48~\ns$ (Full) $\netzero$ pulse, a low-leakage fringe is visible~[\cref{fig:crosshair_sim}(d)] corresponding to the interference condition.

\begin{figure}
    \centering
    \includegraphics{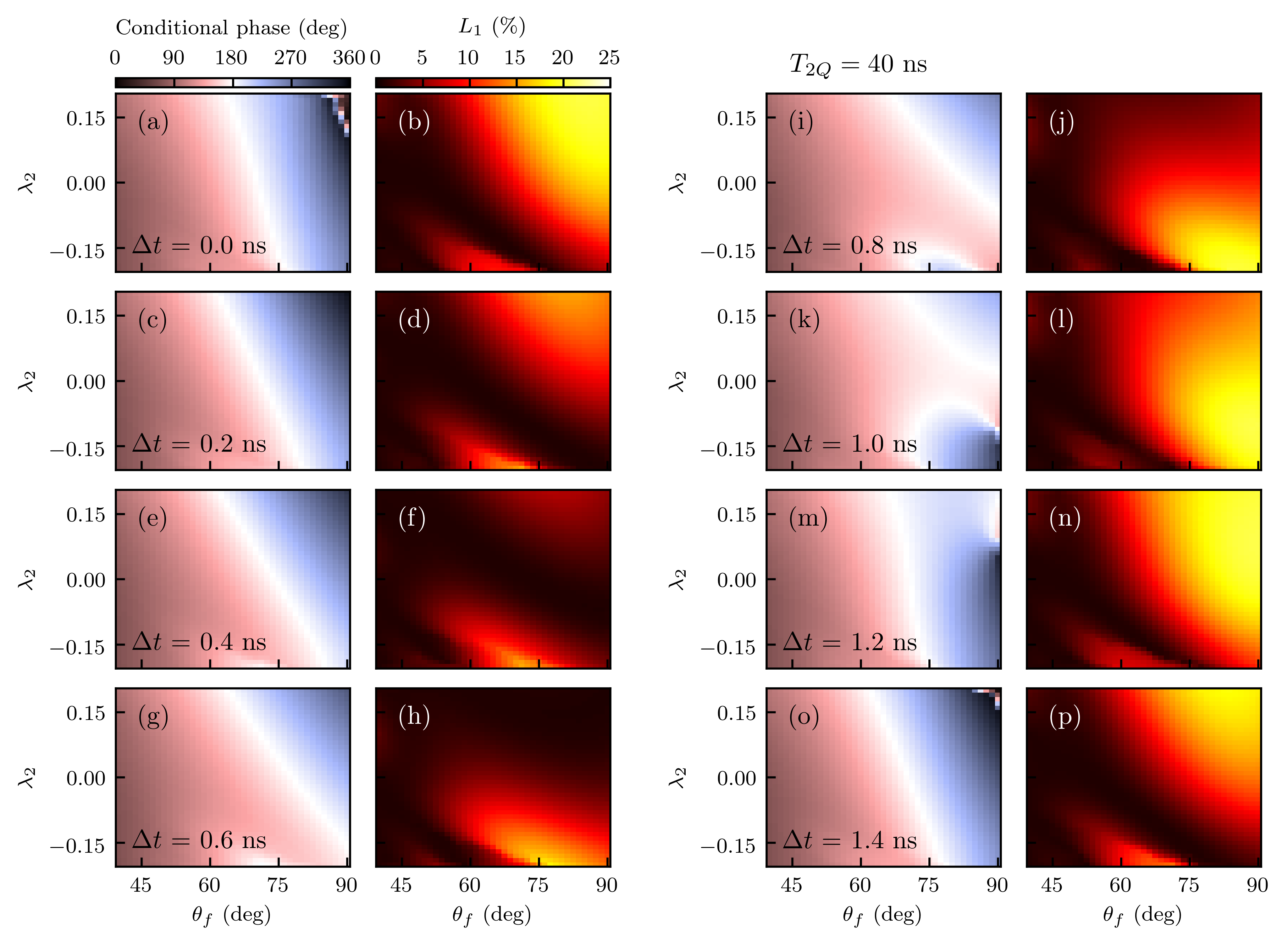}
    \caption{\label{fig:moving_fringe} Moving interference fringes.
    To observe the effect of changing the length of the arms of the interferometer, a buffer ($\Delta t$) is added between the first and second part of the strong $\netzero$ pulse ($\TtwoQ=40~\ns+\Delta t$) in simulation.
    The low-leakage fringe can clearly be seen to move over the landscape.
    }
\end{figure}

The position of the interference fringe should depend on the time between the two halves of the pulse.
This can be explored by adding a buffer time between the two halves of the pulse in simulation.
For a $\TtwoQ=40~\ns$ pulse, the fringe can be seen to move over the leakage landscape~(\cref{fig:moving_fringe}).
The period corresponds to the expected period of $\sim 1/800~\MHz=1.25~\ns$.

\subsection{Leakage modification for randomized benchmarking}
\label{sec:leakage_mod_RB}
Leakage out of the computational subspace is determined using the protocol introduced in~\cite{Wood18_SOM}, which constitutes a modification of the randomized benchmarking protocol.

To determine the populations in the ground ($g$), first-excited ($e$), and second-excited ($f$) states we follow the procedure described in~\cite{Asaad16_SOM}.
In this procedure, a given experiment is performed in two different variants: once in the normal way, giving signal $S_\mathrm{I}$, and once with a $\pi$ pulse on the $g-e$ transition appended at the end of the sequence just before the measurement, giving signal $S_\mathrm{X}$.
When the respective reference signals $V_0$, $V_1$, and $V_2$ of a transmon qubit prepared in the $g$, $e$ and $f$ state are known, the respective populations of the $g$ and $e$ states, $P_0$ and $P_1$, can be extracted using
\begin{equation}
\begin{bmatrix}
V_0 -V_2 & V_1- V_2 \\
V_1 -V_2 & V_0- V_2
\end{bmatrix}
\begin{bmatrix}
P_0 \\
P_1
\end{bmatrix}
=
\begin{bmatrix}
S_\mathrm{I}-V_2 \\
S_\mathrm{X}- V_2
\end{bmatrix},
\end{equation}
under the assumption that higher-excited  levels are unpopulated (in other words, $P_0 + P_1 + P_2 = 1$, where $P_2$ is the population in the $f$ state).

Following~\cite{Wood18_SOM}, we fit the population $P_{\compsub}$ in the computational subspace $\compsub$ to a single exponential
\begin{equation}
P_{\compsub} (N_\mathrm{Cl.} )=A+B\lambda_1^{N_\mathrm{Cl.}},
\end{equation}
where $N_\mathrm{Cl.}$ is the number of Cliffords.
The average leakage (${\leak}$) and seepage ($\seep$) rates~[\cref{eq:leakage,eq:seepage}] per Clifford can then be estimated as
\begin{equation}
{\leakClifford}=(1-A)(1-\lambda_1 ),
\end{equation}
\begin{equation}
\seepClifford =A(1-\lambda_1 ).
\end{equation}
Using the fitted value of $\lambda_1$, the survival probability $M_0$ is then fitted to a double exponential of the form
\begin{equation}
M_0 (N_\mathrm{Cl.})= A_0+B_0 \lambda_1^{N_\mathrm{Cl.}}+C_0 \lambda_2^{N_\mathrm{Cl.}}.
\end{equation}
The average gate infidelity per Clifford $\infidClifford$ is given by
\begin{equation}
\infidClifford =1-\frac{1}{d_1}  \left[ (d_1-1) \lambda_2 + 1-{\leak}\right],
\end{equation}
with $d_1=\dim\compsub$.
We note that if the leakage is weak ($\lambda_1 \ll \lambda_2$  and $B \ll A$), this reduces to the conventional randomized benchmarking formula.
We refer to this experiment as the reference sequence.

This method is used in combination with interleaved randomized benchmarking~\cite{Magesan12b_SOM} to extract the average gate infidelity ($\infidCZ$) and leakage ($\leakCZ$) per CZ gate
\begin{equation}
\label{eq:interleaved}
\infidCZ=1-\frac{1-\infidInterleaved}{1-\infidClifford },
\end{equation}
\begin{equation}
\label{eq:leak_interleaved}
\leakCZ=1-\frac{1-\leakInterleaved}{1-\leakClifford },
\end{equation}
where $\infidInterleaved$ ($\leakInterleaved$) stands for the average gate fidelity (leakage) in the interleaved sequence of the interleaved randomized benchmarking experiment.

\end{document}